\documentclass[twocolumn,showpacs,amsmath,amssymb,10pt, aps]{revtex4-2}
\usepackage{graphicx,color}
\usepackage{subcaption}
\usepackage{soul}
\usepackage{bbm, bm}
\usepackage{color}
\usepackage{xcolor}

\usepackage{hyperref}
\usepackage[T1]{fontenc}
\usepackage[cp1250]{inputenc}
\usepackage{amsfonts}
\usepackage{amssymb, amsbsy}
\usepackage{amsmath, bbm}
\usepackage{epstopdf}
\usepackage{physics}

\usepackage{ulem}
\usepackage[nameinlink]{cleveref}
\begin{document}
\title{Optimization of two-photon excitation by indistinguishable photons \\ in a three-level atom}
%\title{Optimal two-photon excitation of a three-level atom by indistinguishable photons }

\author{Masood Valipour$^{1,2*}$, Gniewomir Sarbicki$^{1,2}$, Karolina S\l{}owik$^{1,2}$, and Anita D\k{a}browska$^3\dagger$}

\affiliation{$^1$Institute of Physics, Faculty of Physics, Astronomy and Informatics,  Nicolaus Copernicus University, Grudzi\c{a}dzka 5/7, 87--100 Toru\'n, Poland\\ $^2$ Institute of Advanced Studies, Nicolaus Copernicus University, Wile\'{n}ska 4, 87-100 Toru\'{n}, Poland
\\$^3$Institute of Theoretical Physics and Astrophysics, Faculty of Mathematics, Physics and Informatics, University of Gda\'nsk, Wita Stwosza 57, 80--308 Gda\'nsk, Poland\\$^*$masood.valipour@umk.pl, $^{\dagger}$anita.dabrowska@ug.edu.pl}

\begin{abstract}

We investigate the excitation of a three-level ladder-type atom by a unidirectional field with a pair of indistinguishable photons. 
Starting from an analytical expression for the two-photon absorption probability, we determine the two-photon state that maximizes the population of the upper atomic state at a 
chosen time and show that in the limit of an infinitely long pulse, perfect excitation is possible. 
The optimal state is identified as the time-reversed counterpart of the two-photon state emitted in spontaneous cascade decay. 
We then compare this ideal excitation strategy with experimentally accessible families of states, including symmetrized Gaussian product states, temporally correlated Gaussian states, and coherent pulses. 
We analyze how the optimal excitation conditions depend on the ratio of atomic decay rates and on the separation of the atomic transition frequencies. 
For indistinguishable photons, quantum interference may shift the maxima of the marginal spectral distribution away from the atomic resonances and qualitatively modify the optimal excitation strategy. 
Our results clarify the role of indistinguishability and correlations in two-photon absorption and provide guidance for designing realistic excitation schemes in quantum-optical light-matter interfaces. 
\end{abstract}

\maketitle

\section{Introduction and motivation}
Two-photon absorption is one of the fundamental nonlinear processes in light-matter interaction \cite{GoeppertMayer1931,KaiserGarrett1961,GeaBanacloche1989,Javanainen1990,RumiPerry2010}, 
which are typically most prominent in strong electromagnetic fields \cite{alexanian1999}. 
In a ladder-type quantum system, two-photon absorption enables excitation of an upper level through the cooperative action of two photons whose combined energy matches
 the energy gap between the initial and final states. Since the pioneering studies of multiphoton transitions \cite{GoeppertMayer1931,KaiserGarrett1961}, 
 two-photon absorption has become an important concept in atomic, molecular, and optical physics, with applications ranging from spectroscopy and microscopy to 
 selective excitation and quantum control \cite{Denk1990,Oheim2006,RumiPerry2010,Dudovich2004}. At the same time, its practical use is often limited by the weakness 
 of nonlinear optical processes, which motivates the search for quantum states of light capable of enhancing excitation efficiency under low-intensity conditions \cite{RumiPerry2010,Oheim2006}.

A particularly promising route is to employ nonclassical light. Correlated and entangled photon pairs generated, for example, in spontaneous parametric down-conversion (SPDC)
and related sources offer the possibility of engineering not only the total energy of the photon pair, but also its temporal and spectral
structure \cite{Kwiat1995,Law2000,Grice2001,Kolenderski2009,Eckstein2011,Gajewski2016,Dambal2025}. This has led to extensive studies of how photon correlations affect excitation pathways,
modify resonance conditions, and potentially improve the performance of multilevel excitation schemes compared with classical illumination \cite{Raymer2021,Carnio2021,Tabakaev2021,Li2023,Schlawin2024,Castro2025}. 
Recent advances in the generation and characterization of photon pairs make these questions especially timely \cite{Pollmann2024,Serino2024,Panahiyan2023}.

The efficiency of two-photon excitation is determined not only by the total energy carried by the field, 
but also by how well the spectrotemporal structure of the incoming light matches the intrinsic dynamics of the absorber \cite{Raymer2021,DS25,VSSD25}. 
In a three-level ladder system, this matching depends on the transition frequencies and decay rates of the intermediate and 
final states \cite{GeaBanacloche1989,Javanainen1990,Schlawin2021,DS25,VSSD25}. In our previous work \cite{VSSD25},
 we examined this problem for two-photon light with spectrally and spatially distinguishable photons, where each photon was effectively associated with one atomic transition. 
 That analysis showed that the optimal excitation strategy depends qualitatively on the relation between the lifetimes of the excited states, 
 and that experimentally accessible pulse shapes approximate the optimal behavior only in selected parameter regimes.

In the present work, we address the complementary and physically distinct situation in which the two photons occupying the same spatial mode are spectrally indistinguishable, 
so that each of them can excite either of the two atomic transitions. This changes the structure of the absorption process in an essential way 
because the two excitation pathways become quantum mechanically indistinguishable and interfere. 
Consequently, the problem is not a simple extension of the distinguishable-photon case. 
Instead, it requires understanding how bosonic symmetrization and interference reshape the optimal temporal and spectral properties of the input field,
 and how these effects depend on the atomic parameters.

We describe the interaction between light and matter within the Wigner-Weisskopf approximation \cite{Scully1997}. The evolution of a quantum system driven by a propagating wave packet with a definite photon number \cite{Loudon00} has been formulated within the input-output framework \cite{GarCol85,GardinerZoller10,WisemanMilburn2010}, with important results given in Refs.~\cite{GEPZ98,GJNC12a,BCBC12,Baragiola17}. The reduced dynamics of such systems and the corresponding analytical expressions for quantum trajectories have also been developed using collision-model and stochastic approaches \cite{Dabrowska17,Dabrowska19,DS25}. In particular, Ref.~\cite{DS25} derived analytical formulas for the probability of two-photon absorption in a three-level system driven by two-photon light. This result provides the starting point for our analysis.

Using the formula derived in Ref.~\cite{DS25}, we first characterize the two-photon state that maximizes the population of the upper atomic level at a chosen time. We then compare this ideal excitation scenario with experimentally relevant families of two-photon states and with coherent pulses carrying an average of two photons. In this way, we identify the conditions under which high excitation probability can be achieved with realistic states of light, and we clarify the role played by photon indistinguishability, spectral correlations, and the ratio of the decay rates $\Gamma_e$ and $\Gamma_f$. The case of optimal excitation by spectrally distinguishable photons discussed in Ref.~\cite{VSSD25} serves as a natural reference point throughout our analysis.

The paper has the following structure. \Cref{sec: Model} presents a model of atomic excitation by a unidirectional two-photon field.
\Cref{sec: Properties of the optimal excitation} discusses the properties of the  two-photon state that optimally excites the system. 
\Cref{sec: Excitation Efficiency of Experimentally Accessible Two-Photon States} contain a discussion of the optimal conditions for atomic excitation for two-photon states accessible in the laboratory. 
\Cref{Section: Coherent state} presents the results for the optimal excitation of the system by a coherent pulse.

\section{Model}
\label{sec: Model}

We consider a traveling wave packet in two-photon state interacting with a quantum system. We describe the light-atom interaction within the input-output formalism \cite{GarCol85,GardinerZoller10,WisemanMilburn2010}. We assume that a continuous-mode unidirectional field is prepared in the pure two-photon state of the form
\begin{equation}\label{eq: tps1}
\ket{2_{\Phi}} =\frac{1}{\sqrt{\mathcal{N}}}\int_{t_0}^{+\infty}dt_1\int_{t_0}^{+\infty}dt_2\Phi(t_1,t_2)\hat{a}^{\dagger}(t_1)\hat{a}^{\dagger}(t_{2})\ket{vac},
\end{equation}
where the annihilation operator written in the time domain is defined as 
 \begin{equation}
 \hat{a}(t)=\frac{1}{\sqrt{2\pi}}\int_{-\infty}^{+\infty}{ d}\omega\hat{a}(\omega){e}^{-{i}\omega t}
 \end{equation}
and the field operators $\hat{a}(t)$ and $\hat{a}^{\dagger}(t^{\prime})$ follow commutation relations,
\begin{equation}
[\hat{a}({t}),\hat{a}^{\dagger}({t^{\prime}})]=\delta(t-t^{\prime}),\;\;\;[\hat{a}({t}),\hat{a}({t^{\prime}})]=0.
\end{equation}
The normalization factor in (\ref{eq: tps1}) is given by 
\begin{equation}
\mathcal{N}=\int_{t_0}^{\infty}dt_1\int_{t_0}^{\infty}dt_2\left(\vert\Phi(t_1,t_2)\vert^{2}+\Phi^{\ast}(t_2,t_1)\Phi(t_1,t_2)\right).
\end{equation}
One can check that the maximum value of $\mathcal{N}$ is equal to $ 2\int_{t_0}^{\infty}dt_1\int_{t_0}^{\infty}dt_2\vert\Phi(t_1,t_2)\vert^{2}$ and is reached when $\Phi(t_1,t_2)=\Phi(t_2,t_1)$. Clearly, only the symmetric part of the subintegral function $\Phi(t_1,t_2)$ gives a nonzero contribution to (\ref{eq: tps1}) and thus the formula for the two-photon state can be rewritten in the form
\begin{equation}\label{eq: tps1sym}
\ket{2_{\Phi}} =\int_{t_0}^{\infty}dt_1\int_{t_0}^{\infty}dt_2\Phi_{\rm sym}(t_1,t_2)\hat{a}^{\dagger}(t_1)\hat{a}^{\dagger}(t_{2})\ket{vac},
\end{equation}
where 
\begin{equation}\label{eq: symetrisation}
\Phi_{\rm sym}(t_1,t_2)=\frac{1}{2\sqrt{\mathcal{N}}}
\left(\Phi(t_1,t_2)+\Phi(t_2,t_1)\right).
\end{equation}
The joint temporal distribution is given by
\begin{equation}\label{eq: prob_density_time}
p(t_1,t_2)= 2|\Phi_{\rm sym}(t_1,t_2)|^2 .
\end{equation}
By integrating the joint probability density, 
\begin{equation}
p(t)= \int_{t_{0}}^{\infty}dt\, p(t,t_2)= \int_{t_{0}}^{\infty}dt\,p(t_1,t),
\end{equation} 
we obtain the marginal probability density function in the time domain. The marginal distributions are the same for both photons, clearly in both the time and spectral domains. Let us emphasize that for a two-photon unidirectional field, marginal distributions play a key role in the analysis of the field properties, as they are directly related to the measurements that can be performed on the field. One can verify that $p(t)=0.5f(t)$, where $f(t)=\langle a^{\dagger}(t)a(t)\rangle$ is the mean flux of the light beam in units of photons per unit of time.  For the marginal distribution in the frequency domain, represented by $p(\omega)$, we have $p(\omega)=0.5\tilde{f}(\omega)$, where $\tilde{f}(\omega)=  \langle a^{\dagger}(\omega)a(\omega)\rangle$ is the mean number of photons in the interval from $\omega$ to $\omega+d\omega$.

We assume that the two-photon light prepared in the state (\ref{eq: tps1}) excites a three-level atom in a ladder-type configuration with nondegenerate states, denoted, respectively, by $\ket{g}$, $\ket{e}$, and $\ket{f}$. The Hamiltonian of the atom has the form
 \begin{equation}\label{eq: Hamiltonian of the system}
 \hat{H}=-\omega_{eg}\ketbra{g}{g}+ \omega_{fe}\ketbra{f}{f},
 \end{equation}
where $\omega_{eg}$ and $\omega_{fe}$ are the atomic transition frequencies. The energy of the state $|e\rangle$ is chosen at zero, and $\hbar=1$ set, for simplicity. We consider the model of light-matter interaction with the following assumptions: the rotating-wave approximation, the flat coupling constant, and the extension of the lower limit of integration over frequency to minus infinity  \cite{Scully1997, GarCol85, Loudon00, GardinerZoller10, WisemanMilburn2010}. The evolution operator of the system composed of the atom and the unidirectional field, in the interaction picture that eliminates the free evolution of the field, has the form
 \cite{GardinerZoller10}
 \begin{equation}
 \hat{U}(t)=\overleftarrow{T}\exp\left\{\int_{t_{0}}^{t}\!\left(-i\hat{H}+\left(\hat{L} \hat{a}^{\dagger}(s)-\hat{L}^{\dagger}\hat{a}(s)\right)\!\right)\!ds\!\right\},
 \end{equation}
 where $\overleftarrow{T}$ is the time-ordering operator and the coupling operator $\hat{L}$ is defined as
\begin{equation}\label{eq: coupling_operator}
 \hat{L}=\sqrt{\Gamma_{e}}\ketbra{g}{e}+\sqrt{\Gamma_{f}}\ketbra{e}{f},
 \end{equation} 
where $\Gamma_{e}$, $\Gamma_{f}$ are the inverse of the lifetime of the excited states of the atom due to spontaneous emission into the spatial mode under consideration. The form of $\hat{L}$ indicates that both photons can excite both atomic transitions, i.e., they carry energies lying close to $\omega_{eg}$ and $\omega_{fe}$. The remaining modes that couple to the system contribute only to losses, as they are in the vacuum state. The influence of the interaction with these modes is neglected, noting that it can be easily included in the discussion if needed.

We assume that interaction between the systems starts at time $t_0$ and that the initial state of the composite system is a pure product state,
 \begin{equation}
 \ket{2_{\Phi}}\otimes \ket{g}.
 \end{equation}

As shown in Refs.~\cite{GEPZ98, GJNC12a, BCBC12, Dabrowska17, Dabrowska19}, the evolution of a quantum system interacting with an $n$-photon wave packet, within the Wigner-Weisskopf approximation, is governed by a set of coupled differential equations. Due to inherent temporal correlations, the reduced dynamics becomes non-Markovian and cannot be described by a single master equation of the Gorini-Kossakowski-Sudarshan-Lindblad (GKSL) form. Specifically, when the amplitude $\Phi(t_1, t_2)$ does not factorize into a product of two temporal profiles, an infinite hierarchy of equations emerges. For a system interacting with a two-photon field, an analytical solution to this hierarchy was derived using a stochastic approach in Ref.~\cite{DS25}. Within this framework, the quantum-trajectory formalism allows for the determination of the probability that a three-level ladder atom, initially in the ground state and driven by two-photon light in state (\ref{eq: tps1}), is found in state $\ket{f}$ at time $t$. The corresponding expression reads \cite{DS25}
\begin{widetext}
\begin{equation}\label{eq: prob_tpa}
P_{f}(t)= \frac{\Gamma_{e}\Gamma_{f}}{\mathcal{N}} \left|\int_{t
_0}^{t}dt_2e^{\left(i\omega_{fe}t_2-\frac{1}{2}\Gamma_{f} (t-t_2)\right)}\int_{t_0}^{t_2}dt_1 e^{\left(i\omega_{eg}t_1 -\frac{\Gamma_{e}}{2}(t_2-t_1)\right)}
\left(\Phi(t_1,t_2)+\Phi(t_2,t_1)\right)\right|^2.
\end{equation}
This formula can be rewritten in the compact form as
\begin{equation}
 P_{f}(t)= 4\Gamma_{e}\Gamma_{f}e^{-\Gamma_{f}t}\left|\int_{t_0}^{t}dt_2e^{\left(i\omega_{fe}+\frac{1}{2}(\Gamma_{f}-\Gamma_{e})\right)t_2}\int_{t_0}^{t_2}dt_1\; e^{\left(i\omega_{eg} +\frac{\Gamma_{e}}{2}\right)t_1}
\Phi_{sym}(t_1,t_2)\right|^2.
\end{equation}
\end{widetext}
The expression for $P_f(t)$ shows a sequence of two successive absorption events, 
the first occurring at time $t_1$ and the second at time $t_2$, while the double 
integration corresponds to summing over all possible realizations of this process.
It is worth emphasizing that the symmetric part of the amplitude
$\Phi(t_1,t_2)$ appears here, reflecting the fact that each photon can excite either of the two atomic transitions. Note that relaxation corresponding to the coupling of the atomic system with additional vacuum modes of its photonic environment can be easily incorporated by adding respective decaying exponential factors to the formula above.  At the onset of the interaction between the atom and the two-photon field,
$P_{f}(t_0)=0$ and $P_{f}(t)$ tends to zero as time approaches infinity. 
At any time $t$, $P_{f}(t)\leq 1$; see Appendix \ref{Appendix: Properties}. With a recipe for $P_{f}(t)$, one can also calculate the mean residence time in state $\ket{f}$ that is given as
\begin{equation}\label{eq: mean_time_excitation}
\overline{T}= \int_{t_0}^{+\infty}dtP_{f}(t). 
\end{equation}
 This quantity provides an additional characterization of the excitation process and will be used below to characterize the optimal excitation scenario.

We now proceed to describe the state that optimally excites the atomic system, i.e. maximizes the probability $P_f(t)$ of excitation of the state $|f\rangle$. We will show that the optimal state ensures perfect excitation of the three-level system at a predefined moment $t^\star$. 
Later, we maximize the excitation probability for some standard families of states that are popular in experimental realizations.
Our focus is the value of the probability of atomic excitation maximized over times,
\begin{equation}
\underset{t\in \mathbbm{R}}{\rm{max}} P_{f}(t),
\end{equation}
and the parameters of the light state. 

When examining the aspects of the conditions for optimal excitation of the system by two-photon light, we will discuss the significance of the relation between the decay rates $\Gamma_e$ and $\Gamma_f$, as well as the difference between the atomic transitions,
\begin{equation}
\delta_a=\omega_{fe}-\omega_{eg}.
\end{equation}
As a reference scale, we use the lifetime 
$\Gamma_{f}^{-1}$ of the state $\ket{f}$. We analyze the excitation process
for a broad range of values of the ratio $\Gamma_e/\Gamma_f$. 
The two limiting regimes,
$\Gamma_e \ll \Gamma_f$ and $\Gamma_e \gg \Gamma_f$, provide useful reference points for understanding
the excitation dynamics and the transition between
single-photon-resonance-dominated and two-photon-resonance-dominated excitation scenarios. In physical systems, the two decay rates $\Gamma_e$ and $\Gamma_f$ may differ substantially. For electric-dipole transitions, the spontaneous-emission rate scales as $\Gamma \propto \omega^3 |\mathbf{d}|^2$, where $\omega$ is the transition frequency and $\mathbf{d}$ is the transition dipole moment. Thus, even transitions with comparable frequencies may have very different lifetimes if the corresponding dipole moments differ. This situation can occur in atoms, molecules, and engineered quantum emitters. In molecular systems, in particular, weakly allowed transitions may arise when a transition that is forbidden in a higher-symmetry limit acquires a small but finite dipole moment due to symmetry breaking, vibronic mixing, external fields, or an asymmetric environment \cite{szakacs2021}. The regimes $\Gamma_e \ll \Gamma_f$ and $\Gamma_e \gg \Gamma_f$ therefore represent physically relevant limits of strongly unequal oscillator strengths, while the case $\Gamma_e \sim \Gamma_f$ corresponds to transitions of comparable strength.

\section{Properties of the optimal excitation} \label{sec: Properties of the optimal excitation}

In this section, we describe a two-photon state that maximizes the probability of occupation of state $\ket{f}$ at a chosen time $t^{\star}$. Let us first consider the case in which the duration of the two-photon pulse is finite. As shown in \cite{DS25}, the maximum attainable value of the probability $P_{f}(t^{\star})$ at time
 $t^{\star}>t_{0}$ reads then as follows:
\begin{equation}\label{P-max1}
P^{\rm max}_{f}(t^{\star}) = 1- \frac{1}{\Gamma_{f}-\Gamma_{e}}\left(\Gamma_{f}e^{-\Gamma_{e}(t^{\star}-t_{0})} - \Gamma_{e}e^{-\Gamma_{f}(t^{\star}-t_{0}) }\right),
\end{equation}
and is realized for the two-photon state defined by symmetrizing the function
\begin{align}\label{eq: opt_temp_amplitude}
 \Phi_{\rm opt}(t_1,t_2)=&\frac{e^{\frac{1}{2}\Gamma_{f} t_2-\frac{1}{2}\Gamma_{e}(t_2-t_{1})-i\omega_{fe}t_2-i\omega_{eg}t_1}}{\sqrt{\mathcal{N}_{\rm opt}}}\nonumber\\
& \times\Theta(t^{\star} - t_2) \Theta(t_2 - t_1) \Theta(t_1 - t_0)
 %\chi_{t_0<t_1<t_2<t^{\star}},
\end{align}
where $\Theta(\cdot)$ denotes the Heaviside step function. The normalization factor has the form
\begin{align}
\mathcal{N}_{\rm opt}=\frac{e^{\Gamma_{f} t^{\star}}}{\Gamma_{e}\Gamma_{f}}\bigg[&1-e^{-\Gamma_{f} (t^{\star}-t_0)}\nonumber\\ &+\frac{\Gamma_{f}}{\Gamma_{e}-\Gamma_{f}}\left(e^{-\Gamma_{e}(t^{\star}-t_0)}-e^{-\Gamma_{f}(t^{\star}-t_0)}\right)\bigg].
\end{align}

\begin{figure}[b]
    \includegraphics[width=7cm]{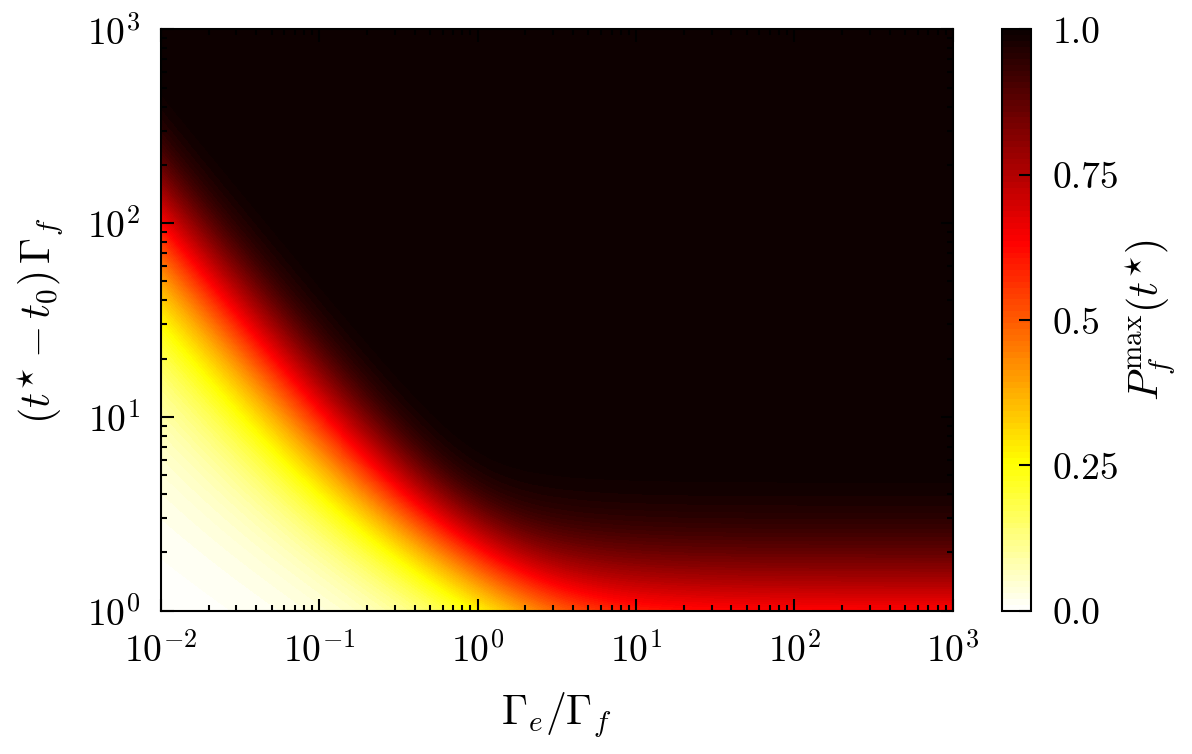}
    \caption{Maximum excitation probability $P_f^{\max}(t^\star)$ for the optimal finite-duration pulse, calculated from Eq. (\ref{P-max1}), as a function of the scaled pulse duration $(t^\star-t_0)\Gamma_f$ and the ratio $\Gamma_e/\Gamma_f$.}
    \label{fig: P_optimum_FiniteTime}
\end{figure}
Note that if $t_{0}\to -\infty$, then the probability $P_{f}(t^{\star})$ approaches unity. Before considering this idealized limit, it is instructive to examine how rapidly the maximum excitation probability approaches unity for finite-duration pulses. Figure \ref{fig: P_optimum_FiniteTime} shows $P_f^{\max}(t^\star)$ given by Eq. (\ref{P-max1}) as a function of the pulse duration $(t^\star-t_0)\Gamma_f$ and the ratio $\Gamma_e/\Gamma_f$. One can see that the convergence towards perfect excitation is rapid. In particular, for $\Gamma_e \gtrsim \Gamma_f$, excitation probabilities close to unity are already obtained for pulse durations of only a few lifetimes, $1/\Gamma_f$. In contrast, for $\Gamma_e \ll \Gamma_f$, substantially longer pulses are required, reflecting the long lifetime of the intermediate state. These results indicate that the limit $t_0\rightarrow -\infty$ serves primarily as a theoretical benchmark. In practice, excitation probabilities very close to unity can be obtained with finite-duration pulses whose length is determined by the larger of the two characteristic decay times.

In the limit $t_{0}\to -\infty$, we obtain the optimal two-photon state of the form
\begin{widetext}
\begin{equation}\label{eq: optstate_temp}
\ket{2_{\rm opt}}=\sqrt{\Gamma_{e}\Gamma_{f}}e^{-\frac{\Gamma_{f} t^{\star}}{2}}\int_{-\infty}^{t^{\star}}dt_2\int_{-\infty}^{t_2}dt_1e^{\frac{1}{2}(\Gamma_{f}-\Gamma_{e})t_2}e^{\frac{\Gamma_{e}}{2}t_1}e^{-i\omega_{fe}t_2-i\omega_{eg}t_1} \hat{a}^{\dagger}(t_2)\hat{a}^{\dagger}(t_1)\ket{vac}. 
\end{equation}
\end{widetext}

It is worthwhile to mention that the two-photon absorption probability can be interpreted as the inner product of the two functions defined, respectively, by the temporal amplitudes of the optimal state (\ref{eq: optstate_temp}) and the two-photon state of the light driving the atom; see Appendix \ref{Appendix: Properties}  for details. The marginal distribution for the optimal state in the time domain is given by
\begin{align}\label{eq: marg_dist_time_opt}
p(t)= &\frac{1}{2}\left\{\tfrac{\Gamma_{f}\Gamma_{e}}{\Gamma_{f} - \Gamma_{e}}\left[e^{-\Gamma_{e}(t^{\star}-t)} - e^{-\Gamma_{f}(t^{\star}-t)}\right]+\Gamma_{f}e^{-\Gamma_{f}(t^{\star}-t)}\right\} 
\end{align}
for $t\leq t^{\star}$ and is equal to zero for $t>t^{\star}$. We would like to note that the formula (\ref{eq: marg_dist_time_opt}) is the arithmetic mean of the marginal distributions obtained for a perfect excitation of three-level system by the distinguishable photons considered in Ref.~\cite{VSSD25}. Rewriting (\ref{eq: marg_dist_time_opt}) in the form 
\begin{align}
   p(t)= \frac{1}{2}\bigg[&
   \frac{\Gamma_{e}}{1 - \frac{\Gamma_{e}}{\Gamma_{f}}}e^{-\Gamma_{e}(t^{\star}-t)} + \frac{\Gamma_{f}}{1 - \frac{\Gamma_{f}}{\Gamma_{e}}}e^{-\Gamma_{f}(t^{\star}-t)}\nonumber\\
    &+\Gamma_{f} e^{-\Gamma_{f}(t^{\star}-t)} \bigg],
\end{align}
allows one to easily recognize the two limit behaviors,
\begin{align*}
    & p(t)\xrightarrow[]{} \frac{1}{2}\left[\Gamma_{e} e^{-\Gamma_{e}(t^{\star}-t)}+\Gamma_{f} e^{-\Gamma_{f}(t^{\star}-t)}\right]  & \text{for} \ & 
    {\Gamma_{e}}\ll{\Gamma_{f}},\\
    & p(t)\xrightarrow[]{} \Gamma_{f} e^{-\Gamma_{f}(t^{\star}-t)}   
& \text{for} \ & {\Gamma_{e}}\gg{\Gamma_{f}}.
\end{align*}

Interestingly, the two-photon state (\ref{eq: optstate_temp}) is the time reversal of the unidirectional two-photon state emitted spontaneously by the three-level atom in a ladder configuration prepared at time $t^{\star}$ in the state $\ket{f}$. One can verify that the optimal state leading to perfect excitation of state $\ket{f}$ at time $t^{\star}$ results in
\begin{equation}
P_{t}(t)=e^{-\Gamma_{f}|t^{\star}-t|}. 
\end{equation}
By referring to Eq. (\ref{eq: mean_time_excitation}), it can be checked that the mean excitation time of state $\ket{f}$ by the optimal state is equal to $2/\Gamma_f$. For a two-photon unidirectional field, one can define the probability distribution of two successive photon counting detections. The formula for the distribution of counting times has then the same form as in the case of distinguishable photons \cite{VSSD25}. The corresponding mean detection times can also be determined. Analogously to Ref. \cite{VSSD25}, the mean time difference between the successive counts is equal to $1/\Gamma_{e}$, i.e., the spontaneous-emission lifetime of state $\ket{e}$. Note that in the case of perfect excitation, this quantity coincides with the difference between the mean times at which the photon absorptions occur.

In the frequency domain, state (\ref{eq: optstate_temp}) has the amplitude of the symmetrized form
\begin{widetext}
\begin{equation}
\tilde{\Phi}_{\rm opt}(\omega_1,\omega_2)=\frac{\sqrt{\Gamma_{e}\Gamma_{f}}e^{i(\omega_1+\omega_2-\omega_{fg})t^{\star}}}{4\pi\big[i\left(\omega_1+\omega_{2}-\omega_{fg}\right)+\frac{\Gamma_{f}}{2}\big]}\left[ \frac{1}{i\left(\omega_1-\omega_{eg}\right)+\frac{\Gamma_{e}}{2}}+\frac{1}{i\left(\omega_2-\omega_{eg}\right)+\frac{\Gamma_{e}}{2}}\right],
\end{equation}
\end{widetext}
where $\omega_{fg}=\omega_{fe}+\omega_{eg}$. The joint probability distribution in the frequency domain is given by
\begin{equation}
\label{Eq: optimum spectral prob}
p(\omega_1,\omega_2)= 2|\tilde{\Phi}_{\rm opt}(\omega_1,\omega_2)|^2 .
\end{equation}
The marginal distribution in the frequency domain then has the form
\begin{widetext}
\begin{equation}\label{eq: optimal_state_marg}
p(\omega)=\frac{\Gamma_{e}(\Gamma_{e}+\Gamma_{f})(\Gamma_{e}+\frac{1}{4}\Gamma_{f})+\Gamma_{e}(\omega_{fe}-\omega_{eg})^2+\Gamma_{f}(\omega-\omega_{eg})^2}{4\pi \left[(\omega-\omega_{eg})^2+\frac{\Gamma_{e}^2}{4}\right]\left[(\omega-\omega_{fe})^2+\frac{(\Gamma_{e}+\Gamma_{f})^2}{4}\right]}
\end{equation}
\end{widetext}
Note that $p(\omega)$ reveals a quantum interference effect; 
consequently, the local extrema of the function (\ref{eq: optimal_state_marg}) are, 
in general, shifted from the atomic transition frequencies $\omega_{eg}$ and $\omega_{fe}$.
Recall that for the optimal state exciting the atom with photons in two spatial modes with non-overlapping frequency distributions, 
the marginal distributions are given by two Lorentzian functions with maxima at 
$\omega_{eg}$ and $\omega_{fe}$ and half widths at half maximum  of
$\Gamma_{e}/2$ and $(\Gamma_{e}+\Gamma_{f})/2$, respectively \cite{VSSD25}. For the function (\ref{eq: optimal_state_marg}), the number and positions of the maxima depend on the transition energy difference $\delta_{a}$ and the decay rates $\Gamma_{e}$ and $\Gamma_{f}$. If $\delta_{a}=0$, then one can check that a single maximum occurs at  $\omega= \omega_{eg}=\omega_{fe}$. When $\delta_{a}\neq 0$, one or two local maxima appear and their positions depends on the ratio $\Gamma_e/\Gamma_f$. For small and moderate values of the ratio $\Gamma_{e}/\Gamma_{f}$, the positions of the maxima are shifted relative to $\omega_{eg}$ and $\omega_{fe}$, with a separation smaller than $|\delta_{a}|$.
For a given $\delta_{a}$, this separation decreases as the ratio $\Gamma_{e}/\Gamma_{f}$ increases, eventually merging for $|\delta_{a}|<\Gamma_{e}$ into one maximum at $\omega = \omega_{fg}/2$. 
When the ratio $\Gamma_{e}/\Gamma_{f}$ tends to zero, then local maxima approach the values $\omega_{eg}$ and $\omega_{fe}$. 
The analytical determination of the local maxima of the function $p(\omega)$ for the limiting cases $\Gamma_{e} \ll \Gamma_{f}$ and $\Gamma_{f} \ll \Gamma_{e}$ can be found in Appendix \ref{Appen: marg_distr_maxima}. A numerical analysis of the locations of local maxima of $p(\omega)$ is given in Ref.~\cite{SupMat1}.

A key role in the discussion of absorption optimization is played by the distribution of the sum of the photon frequencies, $\omega_{+}=\omega_{1}+\omega_{2}$. One can verify that for the optimal state, the probability density of the frequency sum is given by a Lorentzian distribution of the form
\begin{equation}
\label{Eq: opimum marignal plus}
p_{+}(\omega_{+})= \frac{\Gamma_{f}/2}{ \pi \left[(\omega_{+}-\omega_{fg})^2 + \Gamma_{f}^2/4\right]}.
\end{equation}
For details, see Appendix \ref{Appendix: derivation of distribution of sum}.

\begin{figure*}
    \centering
    \includegraphics[width=\textwidth]{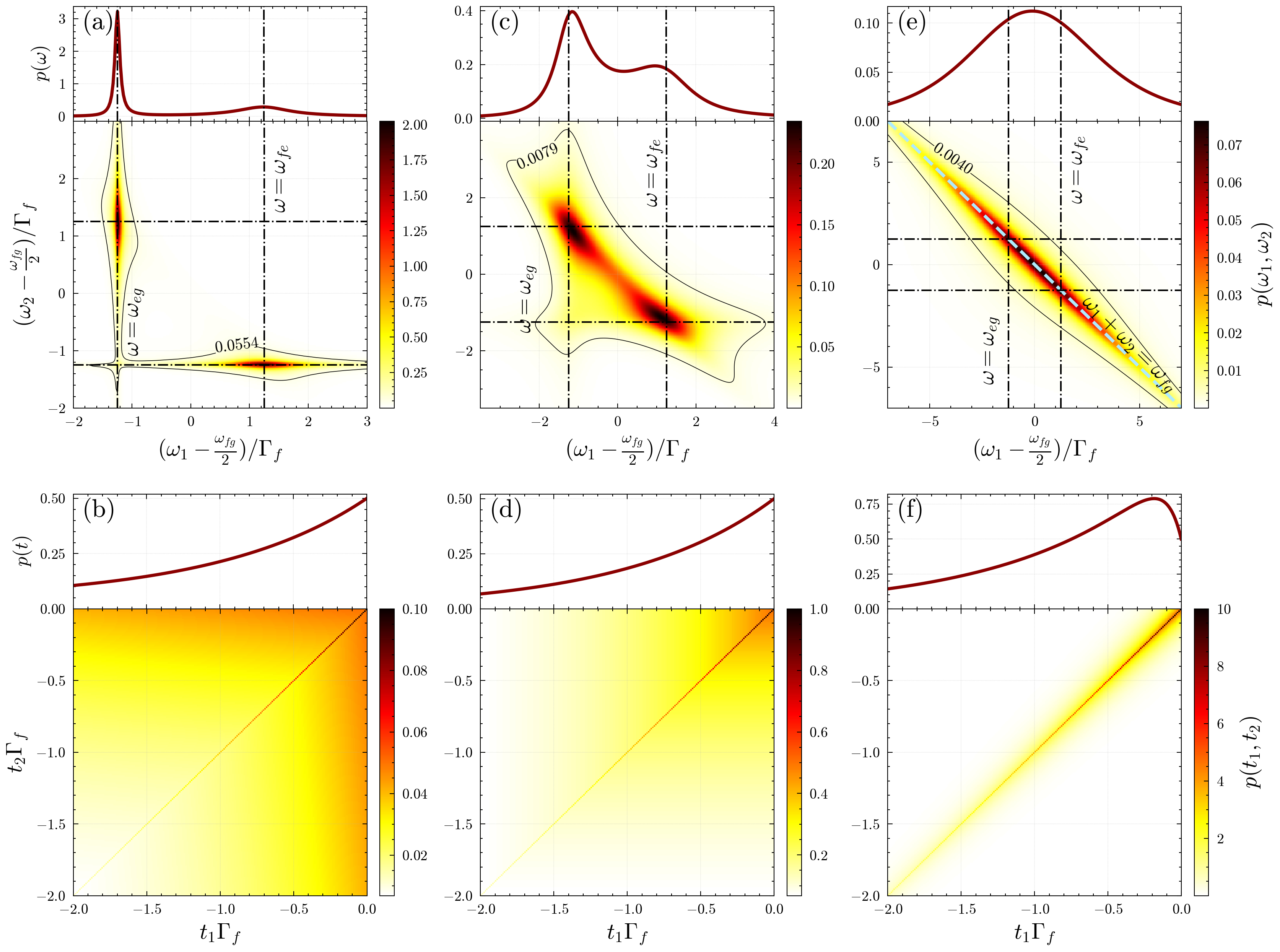}
    \caption{Joint spectral and temporal distributions together with their marginal distributions for the optimal two-photon state $\ket{2_{opt}}$ for the excitation of an atom with the parameter $\delta_{a}=2.5 \, \Gamma_{f}$ and the ratios (a), (b) $\Gamma_e/\Gamma_f = 0.1$, (c), (d) $\Gamma_e/\Gamma_f = 1$, and (e), (f) $\Gamma_e/\Gamma_f = 10$.}
    \label{fig: optimalState_PDF}
\end{figure*}

The joint probability density functions, along with the corresponding marginal distributions in the time and frequency domains for the optimal state, are illustrated in Fig. \ref{fig: optimalState_PDF}. The time-domain distributions are referenced to $t^{\star}=0$. The figure compares three cases with different ratios $\Gamma_e/\Gamma_f$ for a fixed value of $\delta_a/\Gamma_f$. The marginal distributions in the frequency domain clearly display the shift of local maxima resulting from quantum interference. For $\Gamma_e/\Gamma_f=1$, we observe a reduction in the separation between the maxima compared to the $\Gamma_e/\Gamma_f=0.1$ case. As the ratio increases to $10$, the two peaks merge into a single maximum located near $\omega_{fg}/2$. The joint probability density in the frequency domain shows the importance of single and double resonances in the optimal excitation of state $\ket{f}$. Single resonances are defined by the equations $\omega_{1}=\omega_{eg}$ and $\omega_2=\omega_{eg}$. A double resonance is given by the condition $\omega_1+\omega_2=\omega_{fg}$. In the regime $\Gamma_e \ll \Gamma_f$, single resonances contribute significantly, whereas the double resonance dominates for $\Gamma_e \gg \Gamma_f$.
An interactive plot of these distributions is available at the link provided in Ref. \cite{SupMat2}.

In the following sections, we optimize the excitation conditions of an atom for 
experimentally feasible two-photon states. The state corresponding to perfect excitation 
will serve as a reference point for our subsequent discussion.

\section{Excitation Efficiency of Experimentally Accessible Two-Photon States} \label{sec: Excitation Efficiency of Experimentally Accessible Two-Photon States}

In this section, we move to an analysis of photon-pair states described by Gaussian profiles similar to those routinely realized in SPDC experiments \cite{Kwiat1995,Law2000,Grice2001,Kolenderski2009,Eckstein2011,Gajewski2016}. 

\subsection{Result 1}\label{Section: results1}

\begin{figure}[h]
		\includegraphics[width=7cm]{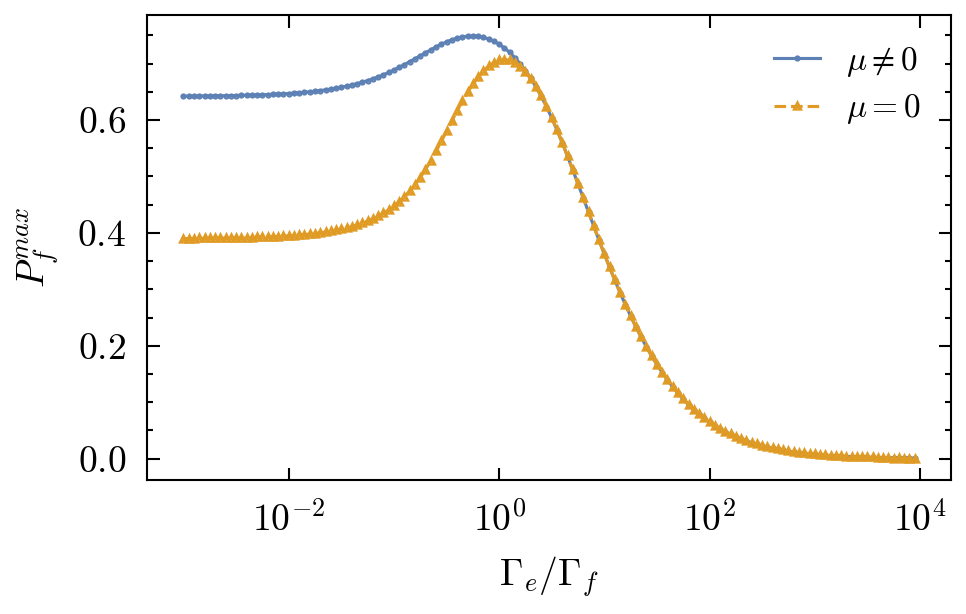}
\caption {Maximum excitation probability of state $\ket{f}$, $P^{max}_{f}$, as a function of the ratio $\Gamma_{e}/\Gamma_{f}$ for the atom with $\delta_{a}=0$ and the input two-photon light defined by (\ref{eq: tps2}) with fixed $\delta_f=0$ satisfying the resonance condition. Results are for the optimized widths of Gaussian profiles with and without optimized value of $\mu$.
}
\label{fig: Pmax_sep_resonance_delta_f=0}
\end{figure}

We analyze the conditions for optimal excitation of the atom by two-photon light defined by a temporal amplitude,
\begin{equation}\label{eq: tps2}
\Phi_{sym}(t_1,t_2)=\frac{1}{2\sqrt{\mathcal{N}}}\left[\psi_{a}(t_1)\psi_{b}(t_2)+\psi_{b}(t_1)\psi_{a}(t_2)\right],
\end{equation}
where $\psi_{a}(t)$ and $\psi_{b}(t)$ are the Gaussian profiles,
\begin{equation}
\psi_{a}(t)=\left(\frac{\Omega_a^2}{2\pi}\right)^{1/4}\exp\left[-\frac{\Omega_a^2}{4}(t-\mu_{a})^2-i\omega_{a}t\right]
\end{equation}
\begin{equation}
\psi_{b}(t)=\left(\frac{\Omega_b^2}{2\pi}\right)^{1/4}\exp\left[-\frac{\Omega_b^2}{4}(t-\mu_{b})^2-i\omega_{b}t\right],
\end{equation}
and 
\begin{equation}
\mathcal{N} = 1+\frac{2\Omega_a\Omega_b}{\Omega_a^2+\Omega_b^2} \exp\left[-\frac{\mu^2\Omega_a^2\Omega_b^2+4\delta_{f}^2}{2(\Omega_a^2+\Omega_b^2)}\right],
\end{equation}
where $\mu = \mu_{b}-\mu_{a}$ and $\delta_{f}=\omega_{b}-\omega_{a}$. 
In the frequency domain, the state has an amplitude of the form
\begin{equation}
\tilde{\Phi}_{sym}(\omega_2, \omega_1)=\frac{1}{2 \sqrt{\mathcal{N}}}(\tilde{\psi}_{a}(\omega_2)\tilde{\psi}_{b}(\omega_1)+\tilde{\psi}_{b}(\omega_2)\tilde{\psi}_{a}(\omega_1)),
\end{equation}
where
\begin{equation}
\tilde{\psi}_{a}(\omega)=\left(\frac{2}{\pi\Omega_{a}^2}\right)^{1/4}\exp\left[-\frac{(\omega-\omega_{a})^2}{\Omega_{a}^2}+i\mu_{a}(\omega-\omega_{a})\right]
\end{equation}
and 
\begin{equation}
\tilde{\psi}_{b}(\omega)=\left(\frac{2}{\pi\Omega_{b}^2}\right)^{1/4}\exp\left[-\frac{(\omega-\omega_{b})^2}{\Omega_{b}^2}+i\mu_{b}(\omega-\omega_{b})\right].
\end{equation}
   When considering Gaussian pulses, we assume that the interaction begins at minus infinity. We are interested in the optimal adjustment of the widths of the Gaussian functions and the positions of their maxima in both the time and frequency domains to the parameters of the atom. 
   
   Analyzing the properties of light under optimized excitation, we will refer to the form of the marginal distributions. In the time domain, the probability density then takes the form
\begin{align}
p(t)=\frac{1}{2\mathcal{N}}&\big\{\left|\psi_{a}(t)\right|^2+\left|\psi_{b}(t)\right|^2+2{\rm{Re}}(\psi_{a}^{\ast}(t)\psi_{b}(t) x)\big\},
\end{align}
and in the frequency domain is given by
\begin{align}
p(\omega)= \frac{1}{2\mathcal{N}} \left\{ \left|\tilde{\psi}_{a}(\omega)\right|^2 + \left|\tilde{\psi}_{b}(\omega)\right|^2 +  2{\rm{Re}}(\tilde{\psi}_{a}^{\ast}(\omega)\tilde{\psi}_{b}(\omega)x)\right\},
\end{align}
where
\begin{equation}
    x=\sqrt{\tfrac{2\Omega_a\Omega_b}{(\Omega_a^2+\Omega_{b}^2)}}\exp\bigg\{-\tfrac{\Omega_a^2\Omega_b^2\mu^2+4\delta_{f}^2-4i\delta_{f}\left(\mu_a\Omega_{a}^2+\mu_{b}\Omega_{b}^2\right)}{4(\Omega_{a}^2+\Omega_{b}^2)}\bigg\}.
\end{equation}

To illustrate the role of the time delay between the Gaussian profiles in the optimal excitation of state $\ket{f}$, we consider an example involving an atom with equidistant energy levels and a field satisfying $\delta_f=0$, resonant with the atomic transitions, i.e., $\omega_{fe}=\omega_{eg}=\omega_a=\omega_b$.  We compare two scenarios: one with a vanishing time delay, $\mu = 0$, and another with an optimally chosen $\mu$. In both cases, we choose the widths of the Gaussian functions so as to maximize the instantaneous probability of excitation to the state $\ket{f}$. As shown in Fig.~\ref{fig: Pmax_sep_resonance_delta_f=0}, the introduction of a time delay significantly improves the excitation conditions for $\Gamma_e/\Gamma_f \lesssim 1$.
The maximum instantaneous population of state $\ket{f}$ obtained in this case is $0.75$. For bidirectional two-photon light in a state given by the product of two Gaussian profiles, the corresponding maximum was $0.64$ \cite{VSSD25}.
The most substantial gain from optimization $\mu$ is observed at low values of the ratio, specifically, when the lifetime of the intermediate state is much longer than that of state $|f\rangle$. In this regime, the difference in excitation probabilities reaches $0.25$. For the optimal choice of $\mu$ in the limit $\Gamma_{e} \ll \Gamma_{f}$, the maximum excitation probability approaches $0.64$. This value corresponds to the square of the maximum excitation probability for a single-photon transition in a two-level system ($0.80^2 = 0.64$) \cite{Scarani11, Banacloche17}. Under these conditions, the three-level-atom excitation process can be interpreted as a sequence of two independent single-photon transitions. In this regime, the product $\mu \Gamma_{e}$ approaches unity, while the pulse parameters satisfy $\Omega_a/\Gamma_{e} \approx 1.46$ and $\Omega_{b}/(\Gamma_{e}+\Gamma_{f}) \approx 1.46$. 
We have verified that the overlap $|\langle\psi_{a}|\psi_{b}\rangle|^2 = |\int_{-\infty}^{+\infty} dt \psi_{a}^{\ast}(t)\psi_{b}(t)|^2$ is approximately $0.006$ for $\Gamma_{e}/\Gamma_{f} = 0.001$, confirming that the process involves almost independent excitations of two two-level subsystems.
A similar behavior was observed for the excitation of a three-level atom by bidirectional two-photon light \cite{VSSD25}.
As the ratio $\Gamma_{e}/\Gamma_{f}$ increases, the significance of the 
time delay between the Gaussian profiles diminishes and the overall efficiency
of the process drops to zero. In the regime where $\Gamma_{e}\gg \Gamma_{f}$, the ratio $\Omega_a/\Omega_b$ and  $|\langle\psi_{a}|\psi_{b}\rangle|^2$ both approach unity.

\begin{figure}[h]
\includegraphics[width=8cm]{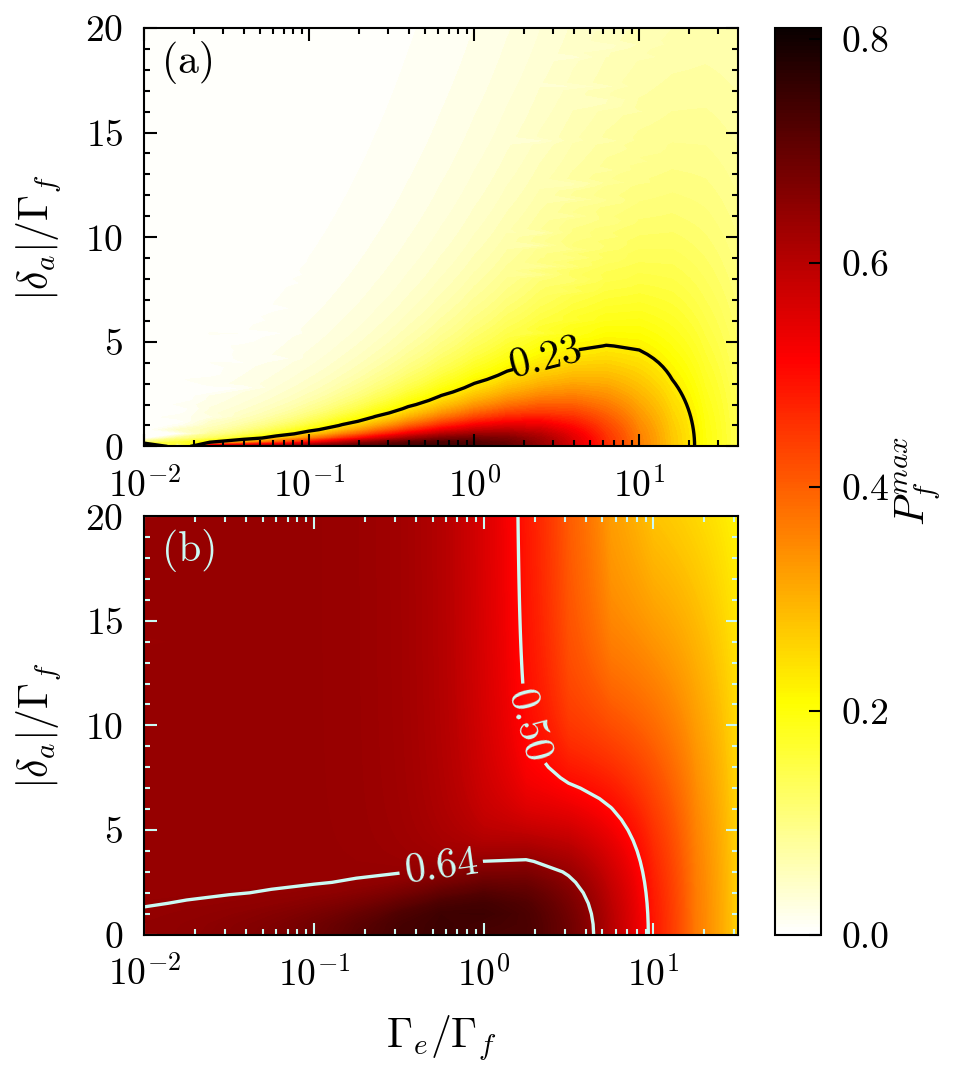}
\caption{Maximum excitation probability of state $\ket{f}$, $P^{max}_{f}$, as a function of the ratio $\Gamma_{e}/\Gamma_{f}$ and $|\delta_{a}|/\Gamma_{f}$ for the input two-photon light defined by (\ref{eq: tps2}). (a) Results for $\delta_f=0$ ($\omega_a=\omega_b=\omega_0$) and $\omega_{0}=\omega_{fg}/2$ with the optimized width of Gaussian profiles and $\mu$. (b)  Results for $\omega_{a}+\omega_{b}=\omega_{fg}$ with the optimized widths of Gaussian profiles and the values of $\mu$ and $\delta_{f}$.
} 
\label{fig: Pmax_sep_out_of_resonace_delta_f=0}
\end{figure}
In the next step, we investigate in detail the role of resonances by considering atoms with different transition energies. We consider a field with $\delta_f=0$, satisfying the two-photon resonance condition.
Figure~\ref{fig: Pmax_sep_out_of_resonace_delta_f=0}(a) illustrates the maximum excitation probability values for a range of $|\delta_a|/\Gamma_{f}$ from $0$ to $20$. As shown in the figure, for small values of the $\Gamma_{e}/\Gamma_{f}$ ratio, efficient excitation of state $|f\rangle$ strictly requires the single-photon resonance condition to be met.
However, as $\Gamma_{e}$ becomes comparable to $\Gamma_{f}$, the field remains capable of efficiently exciting the system over a broad range of $|\delta_a|/\Gamma_{f}$, even when the single-photon resonance condition is violated. We have verified that the introduction of a time delay $\mu$ improves the excitation conditions for $\Gamma_{e}/\Gamma_{f} < 2$ and for detunings $|\delta_a|/\Gamma_{f}$ not exceeding $1.5$. As the $\Gamma_{e}/\Gamma_{f}$ ratio and the detuning $|\delta_a|/\Gamma_{f}$ increase further, the maximum probability $P^{\max}_{f}$ eventually drops to zero.

\begin{figure*}[th!]
\includegraphics[width=\textwidth]{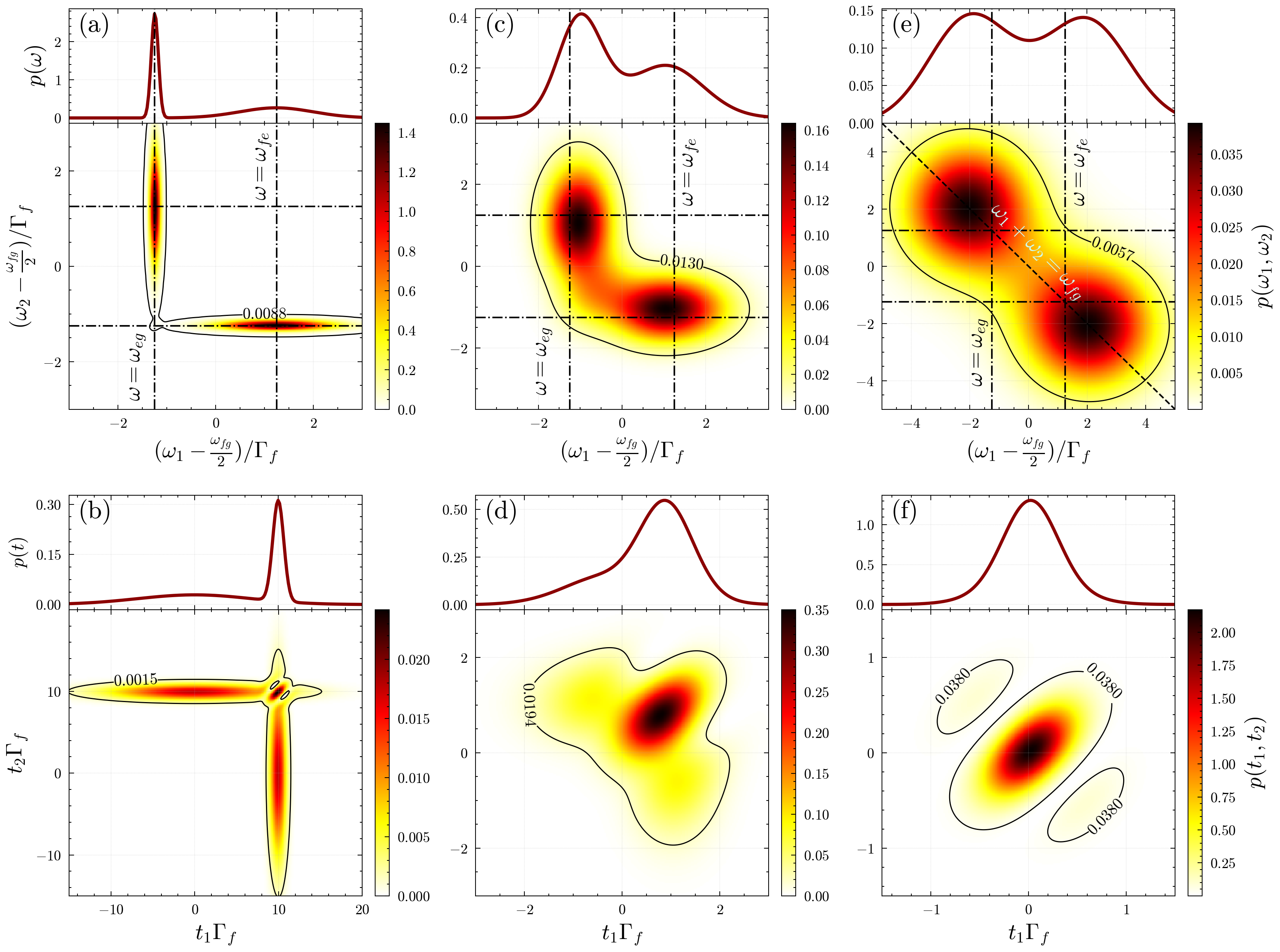}
\caption{Joint spectral and temporal distributions together with their marginal distributions for the two-photon state defined by (\ref{eq: tps2}) obtained for optimal excitation of an atom with the parameter $\delta_{a}=2.5 \, \Gamma_{f}$ and the ratios (a), (b) $\Gamma_e/\Gamma_f = 0.1$, (c), (d) $\Gamma_e/\Gamma_f = 1$, and (e), (f) $\Gamma_e/\Gamma_f = 10$. The corresponding optimized parameters for $\Gamma_e/\Gamma_f = 0.1$ are $\Omega_a=0.15 \, \Gamma_f$, $\Omega_b=1.50 \, \Gamma_f$, $\mu\Gamma_f=9.99$, and $\delta_f=2.49 \, \Gamma_f$;
for $\Gamma_e/\Gamma_f = 1$, they are $\Omega_a=1.02 \, \Gamma_f$, $\Omega_b=1.78 \, \Gamma_f$, $\mu\Gamma_f=1.05$, and $\delta_f=2.09 \, \Gamma_f$;
and for $\Gamma_e/\Gamma_f = 10$, they are $\Omega_a=2.71 \, \Gamma_f$ , $\Omega_b=2.79 \, \Gamma_f$, $\mu\Gamma_f=0.04 $, and $\delta_f=4.16 \, \Gamma_f$.}
\label{fig: UnEnt_PDF}
\end{figure*}

We proceed to study the case when the double resonance condition $\omega_{a}+\omega_{b}=\omega_{fg}$ is satisfied, but $\delta_f\neq 0$ [Fig.~\ref{fig: Pmax_sep_out_of_resonace_delta_f=0}(b)]. 
The probability of populating state $\ket{f}$ is optimized now by properly choosing the parameters $\Omega_a$, $\Omega_b$, $\mu$, and $\delta_{f}$. In general, adjusting $\delta_f$ to the atomic properties improves the excitation probability significantly and in the broad range of parameters. 
For small values of the $\Gamma_{e}/\Gamma_{f}$ ratio and across the entire $|\delta_a|/\Gamma_f$ range that is studied, we observe that the state $\ket{f}$ can be efficiently excited. This requires, however, tuning the field close to resonance, i.e., $\delta_f \approx \delta_a$, and the time shift between Gaussian profiles with values $\mu\Gamma_{e}\approx 1$ (see Supplemental Material \cite{SupMatPDF}). This means that the conditions for double and single resonance are met simultaneously. 
As the inverted lifetimes $\Gamma_{e}$ and $\Gamma_{f}$ become comparable,  
the two-photon excitation probability improves for small $\delta_a\lesssim$ a few $\Gamma_f$. In this case, the optimal two-photon excitation strategy is modified and corresponds to a selection of $\delta_f$ that is different from $\delta_a$. The selection of optimal $\delta_f$ turns out to be a nontrivial function of $\delta_a$ and the $\Gamma_e/\Gamma_f$ ratio (see Supplemental Material \cite{SupMatPDF}, Fig. S2).
Finally, for large $\Gamma_{e}/\Gamma_{f}$ ratios, the atomic excitation with the state (\ref{eq: tps2}) ceases to be efficient. 

The optimized two-photon states for selected atomic parameters are illustrated by the joint probability distributions and the marginal probability distributions in the frequency and time domains in Fig.~\ref{fig: UnEnt_PDF}. Figures~\ref{fig: UnEnt_PDF}(a) and \ref{fig: UnEnt_PDF}(b) correspond to $\Gamma_{e}/\Gamma_{f}=0.1$, Figs.~\ref{fig: UnEnt_PDF}(c) and \ref{fig: UnEnt_PDF}(d) to $\Gamma_{e}/\Gamma_{f}=1$, and Figs.~\ref{fig: UnEnt_PDF}(e) and \ref{fig: UnEnt_PDF}(f) to $\Gamma_{e}/\Gamma_{f}=10$, all with $\delta_{a}/\Gamma_{f} = 2.5$.
For the small ratio $\Gamma_{e}/\Gamma_{f}=0.1$, both the joint probability distribution and the marginal probability distribution in the frequency domain show the important role of single-photon resonances, as seen in Fig.~\ref{fig: UnEnt_PDF}(a). In the time domain, a temporal delay $\mu \neq 0$ between the two Gaussian profiles is clearly visible [Fig.~\ref{fig: UnEnt_PDF}(b)]. This optimized strategy therefore approaches a sequence of two independent single-photon transitions.

A significantly different behavior is observed when $\Gamma_{e}=\Gamma_{f}$ [Figs.~\ref{fig: UnEnt_PDF}(c) and \ref{fig: UnEnt_PDF}(d)]. In this case, the optimized temporal Gaussian profiles of the two photons are centered closer to each other, as indicated by the marginal distribution in the time domain. Nevertheless, the frequency-domain distribution still exhibits two distinct maxima. Note that these maxima are shifted with respect to the atomic transition frequencies indicated by the dot-dashed lines in the figure.

The asymmetry in the marginal distributions, clearly visible in Figs.~\ref{fig: UnEnt_PDF}(a)--\ref{fig: UnEnt_PDF}(d), disappears for $\Gamma_{e}/\Gamma_{f}=10$, as shown in Figs.~\ref{fig: UnEnt_PDF}(e) and \ref{fig: UnEnt_PDF}(f). In this regime, the interference features observed in the joint temporal distribution are associated with a dip in the frequency-domain distribution. The figures clearly demonstrate that the considered two-photon state is well suited to the optimal excitation strategy for state $\ket{f}$ in the regime $\Gamma_e \lesssim\Gamma_f$.
We emphasize that for the large $\Gamma_e/\Gamma_f$ ratio, the efficiency of the scenario exploiting states given in Eq.~(\ref{eq: tps2}) for two-photon excitation is limited. It can be improved with the states described below. 

\subsection{Results 2}\label{Section: results2}

In this section, we consider the two-photon state defined by the symmetrized amplitude $\Phi_\mathrm{sym}(t_1,t_2)$ given by the formula Eq.~(6) with 
\begin{widetext}
\begin{align}\label{eq: tps 3}
\Phi(t_1,t_2)=\sqrt{\frac{\Omega_{+}\Omega_{-}}{2\pi}}\exp\bigg\{-\frac{\Omega_{+}^2}{8}(t_1+t_2-\mu)^2-\frac{\Omega_{-}^2}{8}(t_1-(t_2-\mu))^2-i\omega_{01}t_1-i\omega_{02}t_2\bigg\}.
\end{align}
In the frequency domain, we obtain 
\begin{align}
\tilde{\Phi}(\omega_1,\omega_2)=\sqrt{\frac{2}{\pi\Omega_{+}\Omega_{-}}}\exp\left\{-\frac{\left(\omega_1+\omega_2-(\omega_{01}+\omega_{02})\right)^2}{2\Omega_{+}^2}-\frac{\left(\omega_2-\omega_1-\delta_{f}\right)^2}{2\Omega_{-}^2}+i\mu (\omega_2-\omega_{02})\right\},
\end{align}
where $\delta_{f}=\omega_{02}-\omega_{01}$. 
It can be shown that the marginal distribution in the time domain then takes the form
\begin{align}\label{eq: marg_distribution_time_entangled}
p(t)= &\frac{\Omega}{2\sqrt{\pi}\mathcal{N}}\bigg\{ \exp\left[ -\Omega^{2} (t -\mu)^2\right]  + \exp\left( -\Omega^{2} t^2\right)  \nonumber\\
& +2\exp\left[ -\Omega^{2}\left(t - \frac{\mu}{2}\right)^2 - \frac{\Omega_-^2\mu^2}{4} - \frac{\delta_{f}^2}{\Omega_+^2 + \Omega_-^2}\right]  \cos\left[\frac{2\Omega_+^2\delta_{f}}{\Omega_+^2 +\Omega_-^2}\left(t-\frac{\mu}{2}\right)\right] \bigg\},
\end{align}
\end{widetext}
where
\begin{equation}
\Omega^{2} = \frac{\Omega_+^2\Omega_{-}^2}{\Omega_+^2 + \Omega_-^2}
\end{equation}
and 
\begin{equation}\label{eq: norm_entangled}
\mathcal{N} =1+\exp\left(-\frac{\Omega_{-}^2\mu^2}{4}-\frac{\delta_{f}^2}{\Omega_-^2}\right).
\end{equation}
The temporal marginal distribution is composed of two Gaussian peaks, corresponding to contributions localized around 
$t=\mu$ and $t=0$, and an additional interference term.
The latter is most prominent for $\delta_f=0$ and $\mu=0$, and develops into a fringe pattern as these parameters are nonzero.   
\begin{widetext}
The marginal distribution in the frequency domain is then given by
\begin{equation}
\begin{aligned}
p(\omega)
= 
\frac{1}{\sqrt{\pi\left(\Omega_{+}^2+\Omega_{-}^2\right)}\mathcal{N}}
\Bigg\{
\exp\!\left[
-\frac{4(\omega-\omega_{01})^{2}}{\Omega_{+}^{2}+\Omega_{-}^{2}}
\right]
+
\exp\!\left[
-\frac{4(\omega- \omega_{02})^{2}}{\Omega_{+}^{2}+\Omega_{-}^{2}}
\right]\\[4pt]
%&\qquad
+\,2\exp\!\left[
-\frac{4(\omega- \overline{\omega})^{2}}{\Omega_{+}^{2}+\Omega_{-}^{2}}
-\Omega^2\mu^{2}
-\frac{\delta_{f}^{2}}{\Omega_{-}^{2}}
\right]
\cos\!\left[
\frac{2\Omega_{-}^{2}\mu}{\Omega_{+}^{2}+\Omega_{-}^{2}}\,(\omega -\overline{\omega})
\right]
\Bigg\},
\end{aligned}
 \label{eq:marginal_final}
\end{equation}
\end{widetext}
where 
\begin{equation}
\overline{\omega}= \frac{1}{2}(\omega_{01}+\omega_{02}).
\end{equation}
The marginal probability in frequency consists of two Gaussian peaks at the central frequencies $\omega_{01}$ and $\omega_{02}$, and an additional one centered at their average value. The latter peak is described by an interference term whose fringes vanish at the absence of the time delay $\mu$. Additionally, we derive the probability distribution of the frequency sum, given by 
\begin{equation}
p_{+}(\omega_{+})= \frac{1}{\sqrt{\pi}\Omega_{+}}\exp\left[-\frac{\left(\omega_{+}-2\overline{\omega}\right)^2}{\Omega_{+}^2}\right].
\end{equation}

\begin{figure}[h]
	\includegraphics[width=7cm]{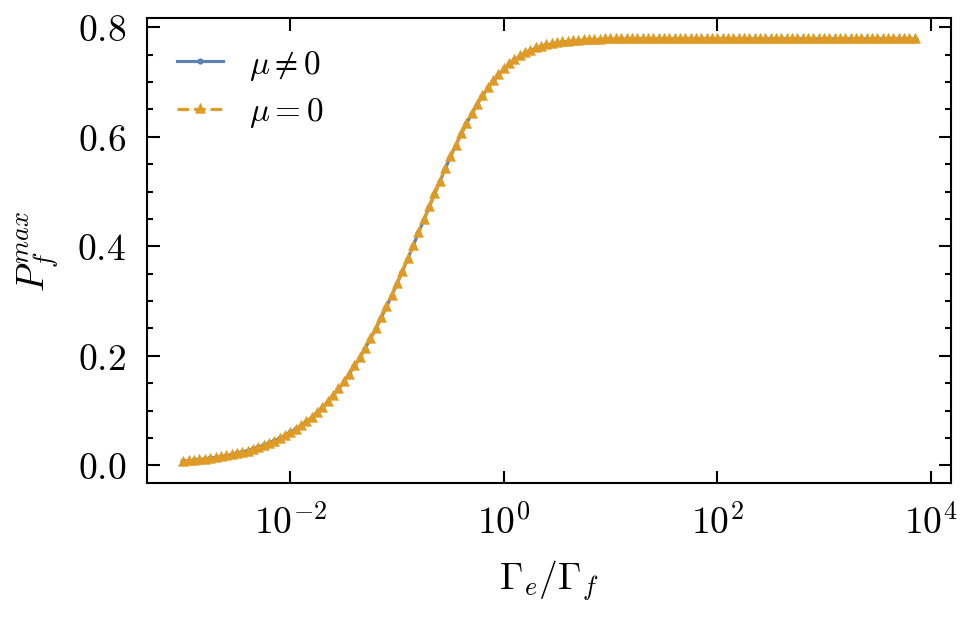}
        \caption{Maximum excitation probability of state $\ket{f}$, $P^{max}_{f}$, as a function of the ratio $\Gamma_{e}/\Gamma_{f}$  for the atom with $\delta_{a}=0$ and for the input two-photon light defined by (\ref{eq: tps 3}) with fixed $\delta_f=0$ satisfying the resonance condition. Results are for the optimized width $\Omega_{\pm}$, with and without optimized value of $\mu$. }
      \label{fig: Corr_Gaussian_GammaRatio}    
\end{figure}

The photonic state defined by (\ref{eq: tps 3}) is now optimized over the widths $\Omega_\pm$, the time delay $\mu$, and the central frequency difference $\delta_f$. As above, we first focus on the case of a resonant field illuminating an atom of equidistant energy levels, i.e., with $\omega_{eg}=\omega_{fe}=\omega_{01}=\omega_{02}$, which means that $\delta_f=0$. In this case, for $\Gamma_e > \Gamma_f$, the maximum excitation probability of state $\ket{f}$ reaches 0.78. In the regime $\Gamma_e \ll \Gamma_f$, the excitation of state $\ket{f}$ by the state (\ref{eq: tps 3}) becomes inefficient and the maximum excitation probability gradually approaches zero.
Contrary to the case described in Sec.~\ref{Section: results1}, the time delay $\mu$ does not notably improve the maximal excitation probability, as shown in Fig.~\ref{fig: Corr_Gaussian_GammaRatio}.

\begin{figure}[t]
\centering
\includegraphics[width=8cm]{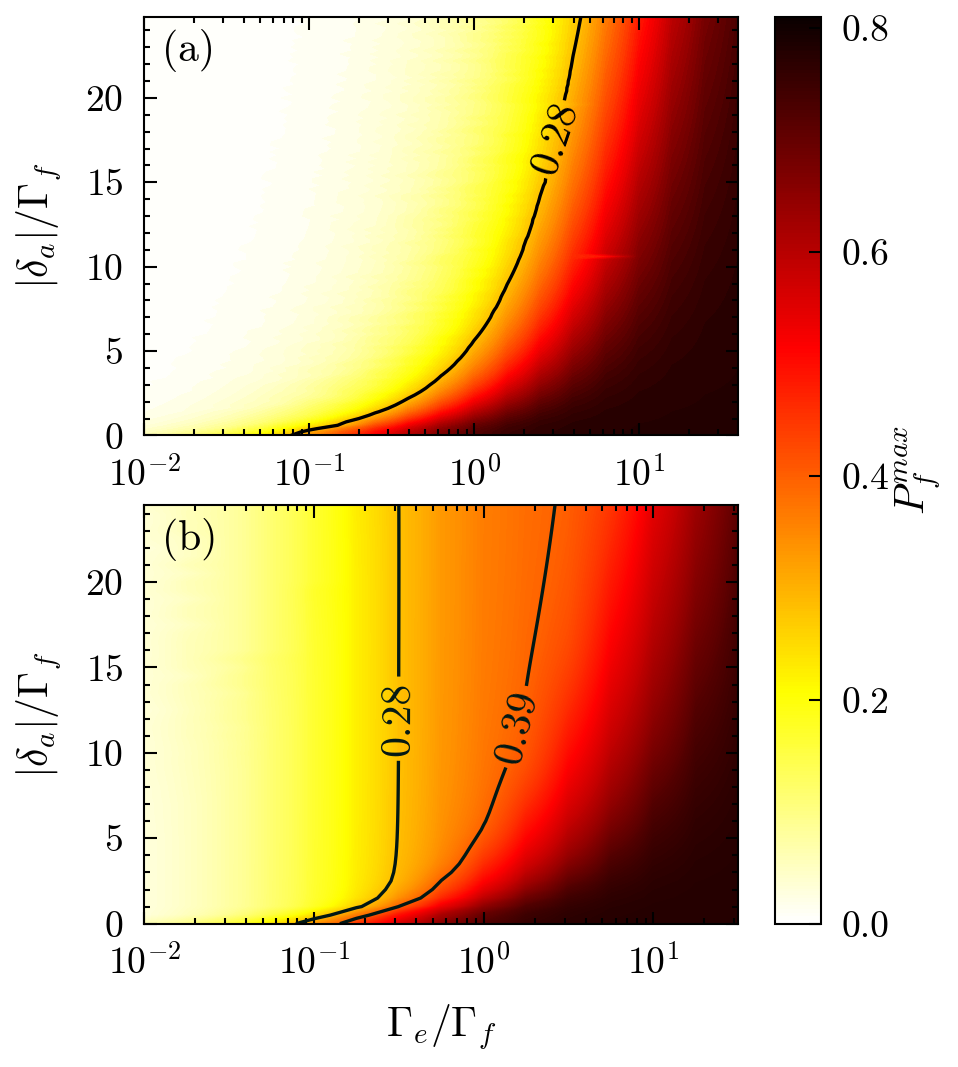}
\caption{Maximum excitation probability $P^{max}_f$, as a function of the ratio $\Gamma_{e}/\Gamma_{f}$ and $|\delta_{a}|/\Gamma_{f}$ for the input two-photon light defined by (\ref{eq: tps 3}). (a) Results for $\delta_f=0$ ($\omega_{01}=\omega_{02}=\omega_0$) and $\omega_{0}=\omega_{fg}/2$ with the optimized widths $\Omega_{\pm}$. (b)  Results for $\omega_{01}+\omega_{02}=\omega_{fg}$ with the optimized widths $\Omega_{\pm}$ and $\delta_{f}$. 
} 

\label{fig: max_prob_corr}
\end{figure}

In the next step, we examine the optimal excitation scenarios for atoms characterized by different transition frequencies ($|\delta_a|\geq 0$), keeping the condition of two-photon resonance $\omega_{01}+\omega_{02}=\omega_{fg} $. 
As shown in Fig.~\ref{fig: max_prob_corr}(a) for the case with $\omega_{01} = \omega_{02}=\omega_0$, considerable excitation probabilities are obtained even for $|\delta_a|>0$, and the range of $|\delta_a|/\Gamma_{f}$ for which the two-photon excitation is efficient grows with the $\Gamma_e/\Gamma_f$ ratio. The maximal probability is reached in the limit of large $\Gamma_e/\Gamma_f$ ratios and is approximately equal to 0.78. 
In the range of small $\Gamma_e/\Gamma_f$ ratios, where the optimal excitation strategy requires a sequence of two single-photon pulses, excitation with the pulse defined by (\ref{eq: tps 3}) is inefficient even if the single- and two-photon resonance conditions are satisfied.
As confirmed by Fig.~\ref{fig: max_prob_corr}(b), the efficiency of the process of excitation grows if the optimization of $\delta_f$ is additionally allowed. For each set of atomic parameters $\delta_a$, $\Gamma_e/\Gamma_f$, the optimal field parameters can be identified (see Supplemental Material \cite{SupMatPDF}, Figs. S3 and S4). In particular, we find that in the limit of $\Gamma_e\gg\Gamma_f$, where the excitation strategy is most efficient, $\Omega_+/\Gamma_f\approx 1.03$ and $\Omega_-/(\Gamma_f+2\Gamma_e)\approx 0.54$. Then, $\delta_f$ approaches zero, which means that $\omega_{01}=\omega_{02}\approx\omega_{fg}/2$.

\begin{figure*}[t]
		\includegraphics[width=\textwidth]{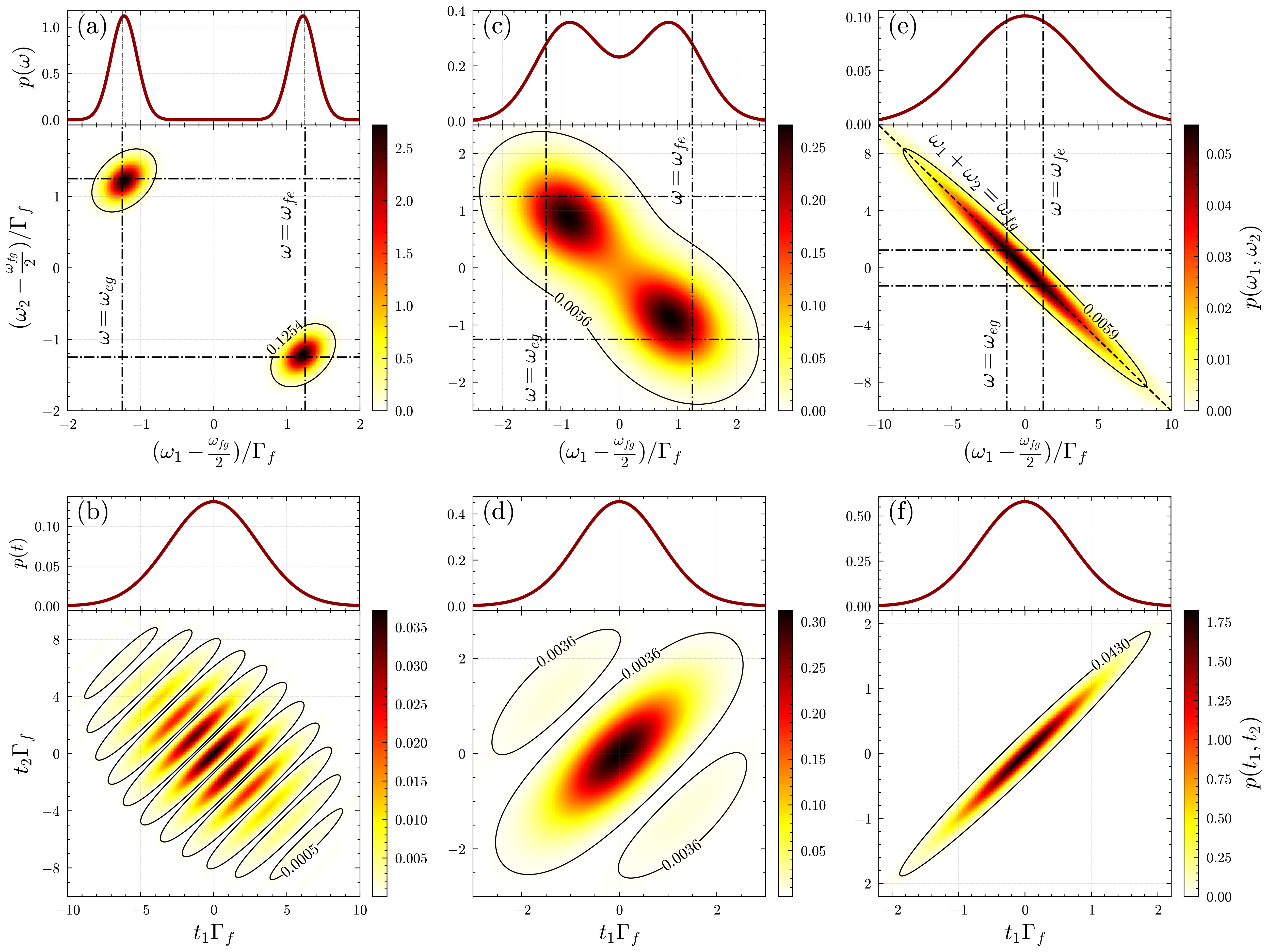}
\caption{Joint spectral and temporal distributions together with their marginal distributions for the two-photon state defined by (\ref{eq: tps 3}) obtained for optimal excitation of an atom with the parameter $\delta_{a}=2.5\,\Gamma_{f}$ and the ratios (a), (b) $\Gamma_e/\Gamma_f = 0.1$, (c), (d) $\Gamma_e/\Gamma_f = 1$, and (e), (f) $\Gamma_e/\Gamma_f = 10$. The corresponding optimized parameters for $\Gamma_e/\Gamma_f = 0.1$ are $\Omega_+=0.42\,\Gamma_f$, $\Omega_-=0.28\,\Gamma_f $,   and $\delta_f=2.45 \,\Gamma_f$;
for $\Gamma_e/\Gamma_f = 1$, they are $\Omega_+=0.89 \, \Gamma_f $ , $\Omega_-=1.21\,\Gamma_f$ ,  and $\delta_f= 1.81 \, \Gamma_f$ ;
and for $\Gamma_e/\Gamma_f = 10$, they are $\Omega_+=1.03\, \Gamma_f$, $\Omega_-=11.09 \,\Gamma_f$, and $\delta_f= 0.002\,\Gamma_f$.}
\label{fig: Ent_PDF}
\end{figure*}

The optimal excitation strategy clarifies based on examination of Fig.~\ref{fig: Ent_PDF}. There, we can distinguish distinct excitation scenarios depending on the atomic properties. 
For low $\Gamma_e/\Gamma_f$ ratios [see Figs.~\ref{fig: Ent_PDF}(a) and \ref{fig: Ent_PDF}(b) where $\Gamma_e/\Gamma_f=0.1$], like before, we find that the optimized illumination corresponds to $\delta_f=\delta_a$. This means that both the single- and two-photon resonance conditions should be fulfilled, leading to a double-peak structure in the marginal frequency distribution.
This results in a prominent interference pattern in the temporal probability distribution shown in Fig.~\ref{fig: Ent_PDF}(b), which, however, vanished in the marginal distribution.

For larger $\Gamma_e/\Gamma_f$ ratios [see Figs.~\ref{fig: Ent_PDF}(c) and \ref{fig: Ent_PDF}(d) for the ratio of 1], the requirement of the single-photon resonance is no longer valid. The optimized strategy is realized by shifting the central frequencies towards their average value. The two peaks in the marginal probability distribution are likewise shifted closer to each other and may spectrally overlap. The interference pattern in the temporal distribution gradually loses its fringes.

Finally, for large $\Gamma_e/\Gamma_f$ ratios [see Figs.~\ref{fig: Ent_PDF}(e) and \ref{fig: Ent_PDF}(f)], we observe that the two peaks merge, leading to a single maximum in the marginal spectral distribution at $\omega=\omega_{fg}/2$. In the time domain, this scenario corresponds to the familiar case of temporally correlated photons \cite{Wasilewski2006,Raymer2021}.

\begin{figure*}
    \centering
    \includegraphics[width=14cm]{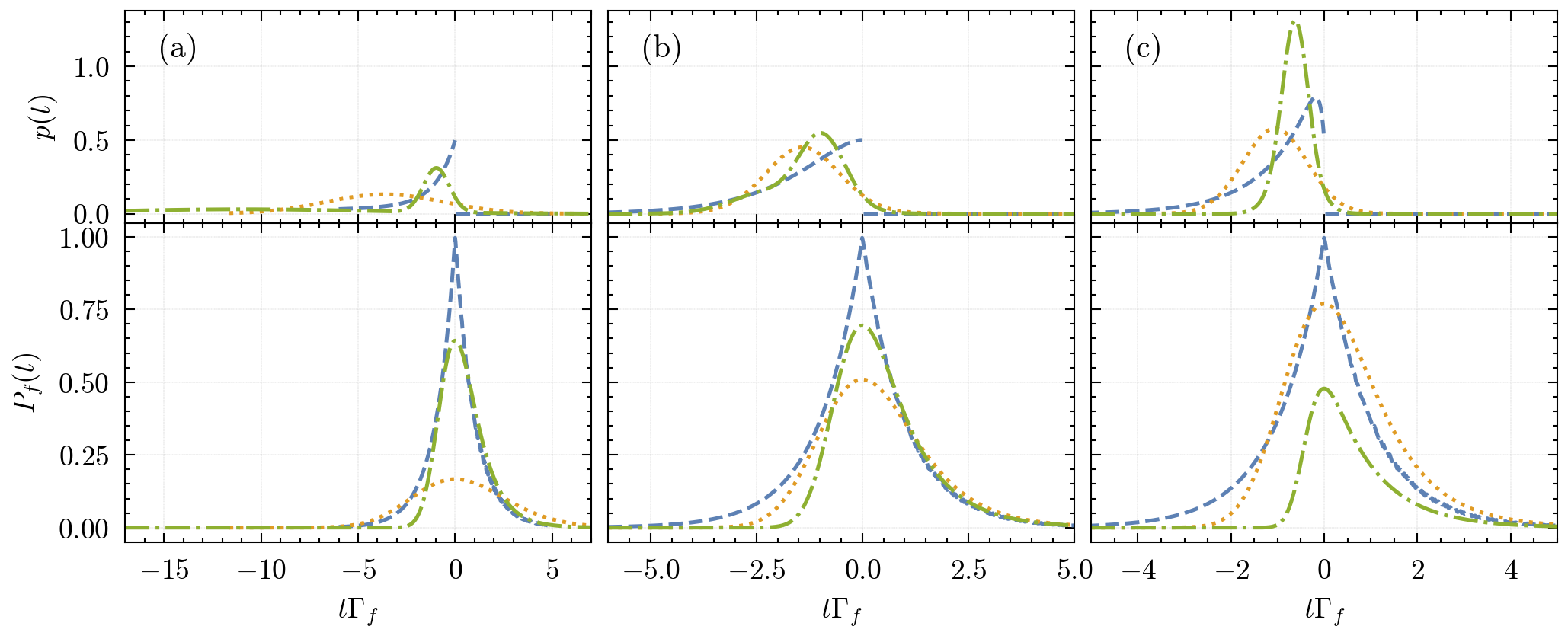}
    \caption{Marginal distribution $p(t)$ (upper panels) and excitation probability from the ground state to the final state, $P_f(t)$ (lower panels), as functions of the scaled time $t\Gamma_f$ for three values of the ratio $\Gamma_e/\Gamma_f= 0.1,\,1,$ and $10$ in (a)--(c), respectively. The blue dashed line corresponds to the optimal state, the green dash-dotted line to the state given by Eq.~(\ref{eq: tps2}),
and the orange dotted line to the state given by Eq.~(\ref{eq: tps 3}). The parameters used in each case are the same as those presented in Figs. 1, 4, and 7, respectively.
}
    \label{fig: dynamics_allCases}
\end{figure*}

In Fig.~\ref{fig: dynamics_allCases}, the optimized Gaussian-pulse excitation strategies for atoms with $\delta_a = 2.5 \Gamma_f$ and ratios $\Gamma_e/\Gamma_f = 0.1$ [Fig.~\ref{fig: dynamics_allCases}(a)], $1$ [Fig.~\ref{fig: dynamics_allCases}(b)], and $10$ [Fig.~\ref{fig: dynamics_allCases}(c)] are compared with the optimal one.
The marginal distributions $p(t)$ for the states are shown in the upper part of the panels. The final-state excitation probabilities $P_f(t)$ are shown in the lower parts.   
For the small $\Gamma_e/\Gamma_f$ ratio 0.1, illumination with the state of the form given by Eq.~\eqref{eq: tps2} outperforms the scenario corresponding to Eq.~\eqref{eq: tps 3}, leading to considerably
larger maximal excitation probability.
This advantage drops for the ratio $\Gamma_e/\Gamma_f = 1$, while the strategy exploiting pulses described by Eq.~\eqref{eq: tps 3} is optimal for large ratios. 

\section{Results for coherent state} \label{Section: Coherent state}

 In this section, we compare the maximum excitation probabilities of state $\ket{f}$ obtained for two-photon light with the corresponding results for excitation of the atomic system by a coherent-state field. To allow for a direct comparison with the results presented in the previous sections, we assume that the coherent field contains, on average, two photons. We determine the optimal parameters for excitation of state $\ket{f}$ for different values of the ratio $\Gamma_e/\Gamma_f$ and different values of $\delta_a$, assuming that the double-resonance condition is satisfied. We consider a coherent pulse defined as \cite{Loudon00}
\begin{equation}\label{eq: coherent_state}
\ket{\{\alpha\}}= \exp\left(\int_{t_0}^{+\infty}(\alpha(t)\hat{a}^{\dagger}(t)-\alpha_j^{\ast}(t)\hat{a}(t))dt\right)\ket{0}, 
\end{equation}
with the amplitude $\alpha(t)$ such that
\begin{equation}
\int_{t_0}^{+\infty}|\alpha(t)|^2dt=\overline{n},
\end{equation}
where $\overline{n}$ is the mean number of photons in the pulse. When performing calculations for a coherent state, it is convenient to introduce a normalized amplitude defined as $\alpha_{0}(t)=\alpha(t)/\sqrt{\overline{n}}$ satisfying the condition 
\begin{equation}
\int_{t_0}^{+\infty}ds|\alpha_{0}(s)|^2=1.
\end{equation}
A coherent pulse is characterized by a single central frequency, which we denote by $\omega_{0}$. The evolution of a quantum system interacting with a unidirectional field in state (\ref{eq: coherent_state}) is given by the master equation in the form \cite{WisemanMilburn2010,Milburn94}
\begin{equation}\label{eq: master_equation}
\dot{\rho}(t)=-i[\hat{H}+\hat{H}_{drive},\rho(t)]+\mathcal{D}[\hat{L}]\rho(t)
\end{equation}
with the driving part of the Hamiltonian given as
\begin{equation}
\hat{H}_{drive}=i\alpha^{\ast}(t)\hat{L}-i\alpha(t)\hat{L}^{\dagger}
\end{equation}
and the damping part as
\begin{eqnarray}
\mathcal{D}[\hat{L}]\rho=\hat{L}\rho \hat{L}^{\dagger}-\frac{1}{2}\{\hat{L}^{\dagger}\hat{L},\rho\}
\end{eqnarray}
with $\{\hat{A},\hat{B}\}=\hat{A}\hat{B}+\hat{B}\hat{A}$, that is associated with the emission of photons to the considered field mode. We assume that the coupling operator $\hat{L}$ is given by (\ref{eq: coupling_operator}). To solve the equation (\ref{eq: master_equation}), it is convenient to work in the rotating frame defined by the unitary transformation,
 \begin{equation}\label{eq: rotating_frame}
 \exp\left\{i\omega_{0}t\ketbra{g}{g}-i\omega_{0}t\ketbra{f}{f}\right\}.
 \end{equation}
Then the Hamiltonian of the atom has the form
 \begin{equation}
 \hat{\tilde{H}}=-\Delta_{1}\ketbra{g}{g}+ \Delta_{2}\ketbra{f}{f}
 \end{equation}
 where $\Delta_1=\omega_{eg}-\omega_{0}$, $\Delta_2=\omega_{fe}-\omega_{0}$. 
 The set of master equations for elements of the density matrix $\rho$ of the atom in the basis $\ket{g}, \ket{e}$ and $\ket{f}$ is provided in Appendix \ref{Appendix: Coherent State}. 
 
We analyze the probability of exciting the atom to the state $\ket{f}$, i.e., the element of density matrix $\rho_{ff}$. To this end, we have numerically solved the set of equations (\ref{eq: master_eq1})--(\ref{eq: master_eq2}) for $\bar{n} = 2$ and the atom prepared initially in the ground state. Below, we present the results of the optimization of atomic excitation for a pulse with a Gaussian profile. We study the case in which the double-resonance condition is satisfied, i.e., $\omega_{0}=\omega_{fg}/2$.  The optimization then concerns the width of the Gaussian function.

\begin{figure}[h]
 \centering
\includegraphics[width=8cm]{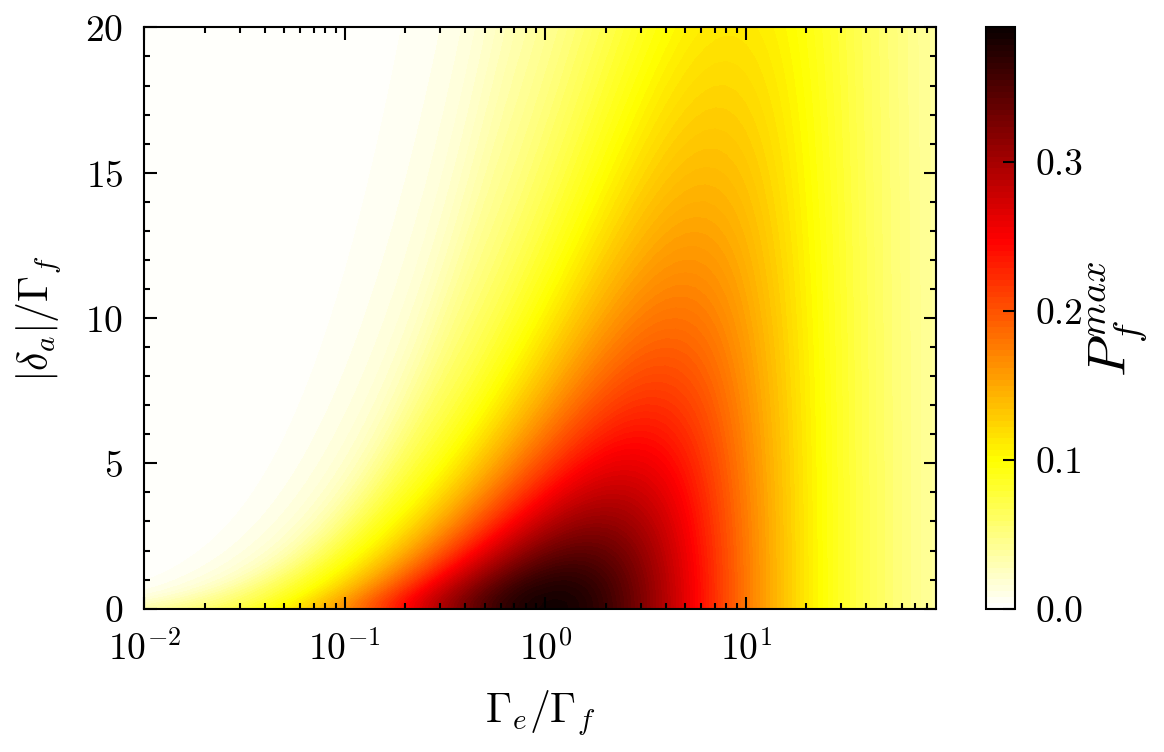}
    \caption{Maximum excitation probability of the state $\ket{f}$, $\rho_{ff}^{max}$, as a function of the ratio $\Gamma_{e}/\Gamma_{f}$ and  $|\delta_a|/\Gamma_{f}$ for a coherent Gaussian pulse with mean photon number $\bar{n} = 2$ satisfying the double-resonance condition, i.e., $\omega_{0}=\omega_{fg}/2$. 
   Results are for the optimized width of Gaussian profile.
   }
\label{fig: Coherent state}
\end{figure}

Figure \ref{fig: Coherent state} presents the maximum value of $\rho_{ff}(t)$ for different values of the ratio $\Gamma_e/\Gamma_f$ and $|\delta_a|/\Gamma_f$. A coherent pulse excites state $|f\rangle$ most effectively in the regime where 
$\Gamma_{e}$ and $\Gamma_{f}$ have comparable values. Note that in this case, while maintaining the double-resonance condition, the state $\ket{f}$ can be effectively excited over a relatively broad range of values of $|\delta_a|/\Gamma_f$. The maximum population of state $|f\rangle$, reaching a value of $0.38$, is obtained for $\Gamma_{e} \approx \Gamma_{f}$ and $\delta_a = 0$. 
In this case, the width of the Gaussian profile is $\Omega = 1.82\Gamma_{f}$. The maximum population of state $|f\rangle$ is lower than that achieved by the two-photon state described in the earlier sections. It remains 
higher than the maximum reached through excitation by the two coherent pulses independently exciting the considered
atomic transitions; see \cite{VSSD25}. In the case of independent coherent excitations 
resonantly tuned to the atomic transitions and carrying an average of one photon each, the maximum population of state $|f\rangle$ 
was $0.23$. This value was obtained for $\Gamma_{e} \ll \Gamma_{f}$ 
by optimally choosing the pulse widths and the delay between their mean 
arrival times of the photons. When the mean arrival times were identical, the maximum
dropped to $0.16$. Thus, a notable difference in atomic excitation between the case of one and two coherent pulses 
is observed when the lifetime of state $|e\rangle$ is significantly longer than 
that of state $|f\rangle$. The use of a coherent pulse coupled to both atomic transitions in the case $\Gamma_e \ll \Gamma_f$ is inefficient since, in this situation, it is not possible to independently match the spectral widths of the two photon excitations to the respective atomic transitions. In both cases, whether using one or two independent coherent pulses, the 
excitation remains ineffective in the regime where $\Gamma_{e}$ significantly 
exceeds $\Gamma_{f}$.  

\begin{figure}[h]
 \centering
\includegraphics[width=7.5cm]{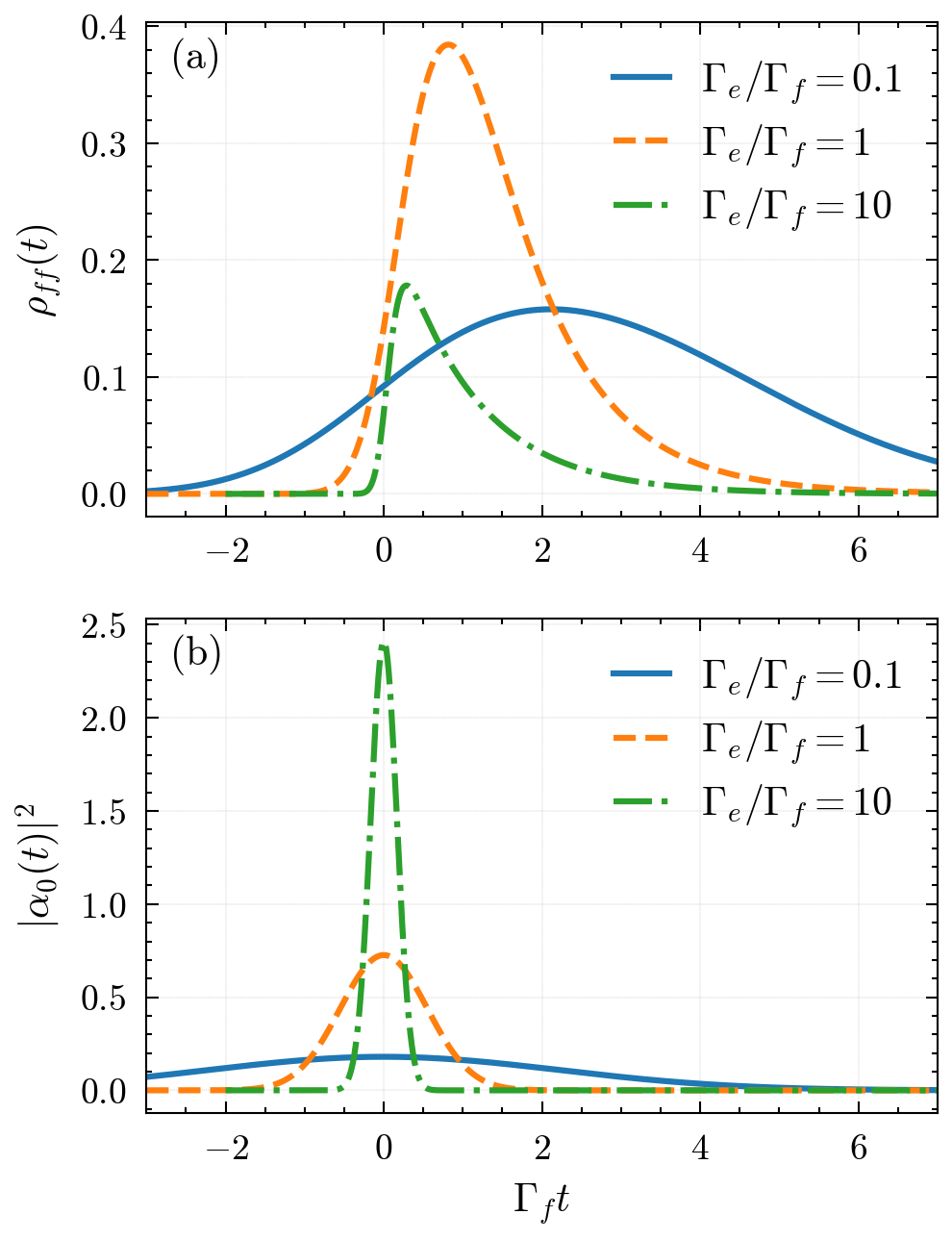}
       \caption{(a) Excitation probability $\rho_{ff}(t)$ of state $\ket{f}$. (b) Squared normalized coherent-state amplitude $|\alpha_0(t)|^2$  for various ratios $\Gamma_e/\Gamma_f$, with mean photon number $\bar{n}=2$ and $\delta_a=0$, and an input coherent Gaussian pulse with the optimal bandwidth. The optimized bandwidths are $0.45, 1.82,$ and $6.05$ for $\Gamma_e/\Gamma_f = 0.1, 1,$ and $10$, respectively. }
\label{fig: Coherent state dynamic}
\end{figure}

Figure \ref{fig: Coherent state dynamic}  shows the excitation of state $\ket{f}$ and the squared modulus of the normalized amplitude as a function of time for three different ratios $\Gamma_e/\Gamma_f = 0.1, 1,$ and $10$, with $\delta_a = 0$ and a field central frequency resonant with the atomic transition frequencies, i.e., $\omega_0 = \omega_{eg} = \omega_{fe}$. In each case, the maximum excitation probabilities are lower than the maximum values obtained by optimizing the parameters of two-photon states. We have checked that while an increase in the average photon number enhances the excitation efficiency, the process remains most effective for the coherent pulse for 
atoms with comparable values of the constants $\Gamma_e$ and $\Gamma_f$.

\section{Summary} \label{sec: Summary}

We investigated optimal excitation of a three-level ladder-type atom by a unidirectional two-photon field composed of spectrally indistinguishable photons. The analysis shows that the excitation process is strongly influenced by quantum interference between indistinguishable excitation pathways, in which each photon can excite either atomic transition. As a consequence, the optimal two-photon state that perfectly excites state $\ket{f}$ differs qualitatively from the case of distinguishable photons and is given by the time-reversed wave packet emitted in spontaneous cascade decay of the atom. A characteristic signature of interference appears in the marginal spectral distribution of the optimal state. For moderate values of the ratio $\Gamma_e/\Gamma_f$, the positions of the local maxima are shifted relative to the atomic transition frequencies and the distance between the peaks becomes smaller than the separation of the corresponding resonances. When $\Gamma_e \gg \Gamma_f$ and $|\delta_a| < \Gamma_e$, the two maxima merge into a single peak located at a half of the two-photon transition frequency. This behavior reflects the role of collective two-photon excitation processes.
The results show that for $\Gamma_e \ll \Gamma_f$, efficient excitation requires matching both single- and double-photon resonances, whereas for $\Gamma_e \gg \Gamma_f$, the excitation dynamics is dominated by the two-photon resonance condition. 

We further demonstrate that experimentally feasible Gaussian states can closely approximate the optimal excitation scenario, while coherent light with the same mean photon number leads to significantly lower excitation probability. These findings show that interference effects determine the spectral structure of the optimal two-photon state and provide a clear physical mechanism responsible for efficient excitation of multilevel quantum systems by nonclassical light. We have also determined how the parameters of the states under consideration depend on the parameters of the atom in the case of the optimal excitation scenario. Deviating from the optimal values results in destructive interference, thereby reducing the probability of exciting state $\ket{f}$. These results provide further insights into how interference effects 
influence the process of two-photon excitation and clarify how photon indistinguishability can be exploited to enhance nonlinear light-matter interactions.

\section*{Acknowledgments}
This project was funded within the QuantERA II Programme that has received funding from the European Union's Horizon 2020 research and innovation programme under Grant Agreement No. 101017733, and with The National Centre for Research and Development, Poland, Grant No. QUANTERAII/1/21/E2TPA/2023.

\section*{Data Availability}
The data that support the findings of this article are openly available \cite{DataGithub2026}.

\nocite{SupMatPDF}
\nocite{apsrev42Control}

\bibliographystyle{apsrev4-2}
\bibliography{TPA_indistinguishable}

\appendix

\section{Properties of $P_{f}(t)$}\label{Appendix: Properties}

The formula for the two-photon state that maximizes the probability $P_f$ at a chosen time $t$ in the general case, i.e., for an arbitrary time $t_0$, was given in Ref.~\cite{VSSD25}. Here, we present the proof only for the case in which $t_0$ tends to infinity. Let us notice that the formula for the probability $P_{f}(t)$
 can then be expressed in terms of a functional,
\begin{align}
\mathcal I[\Phi_{\rm sym}]
=
\int_{-\infty}^{t} ds^{\prime}
e^{\left(i\omega_{fe}+\frac12(\Gamma_f-\Gamma_e)\right)s'}
\nonumber\\
\times\int_{-\infty}^{s^{\prime}} ds
e^{\left(i\omega_{eg}+\frac{\Gamma_e}{2}\right)s}
\Phi_{\rm sym}(s^{\prime},s),
\label{eq:functionalI}
\end{align}
where $\Phi_{\rm sym}(s^{\prime},s)$ is a symmetric two-photon temporal wave function,
\begin{equation}
\Phi_{\rm sym}(s',s)=\Phi_{\rm sym}(s,s'),
\end{equation}
normalized according to
\begin{equation}
2\int_{-\infty}^{\infty} ds'\int_{-\infty}^{\infty} ds\,
\left|\Phi_{\rm sym}(s^{\prime},s)\right|^2 = 1.
\label{eq:norm}
\end{equation}
Thus, we have $P_{f}(t)=\Gamma_{e}\Gamma_{f}e^{-\Gamma_{f}t}\left|I[\Phi_{\rm sym}]\right|^2$. Let us recall that all parameters $\Gamma_e$, $\Gamma_f$, $\omega_{eg}$, and $\omega_{fe}$ are assumed to be positive. Introducing the kernel
\begin{equation}
K_t(s^{\prime},s)
=
e^{A s^{\prime}} e^{B s}
\Theta(t-s^{\prime})\Theta(s^{\prime}-s),
\end{equation}
with
\begin{equation}
A=i\omega_{fe}+\frac{1}{2}(\Gamma_f-\Gamma_e),
\qquad
B=i\omega_{eg}+\frac{\Gamma_e}{2},
\end{equation}
the functional (\ref{eq:functionalI}) can be written in the form
\begin{equation}
\mathcal I[\Phi_{\rm sym}]
=
\iint_{\mathbb R^2} ds^{\prime}ds\,
K_t(s^{\prime},s)\,\Phi_{\rm sym}(s^{\prime},s).
\end{equation}
Since $\Phi_{\rm sym}$ belongs to the subspace of symmetric functions, only the symmetric part of the kernel contributes,
\begin{equation}
K_t^{(\mathrm{sym})}(s^{\prime},s)
=
\frac{1}{2}
\left[
K_t(s^{\prime},s)+K_t(s,s^{\prime})
\right].
\end{equation}
Applying the Cauchy-Schwarz inequality in this subspace yields
\begin{equation}
\left|\mathcal I[\Phi_{\rm sym}]\right|
\le
\left\|K_t^{(\mathrm{sym})}\right\|_2
\left\|\Phi_{\rm sym}\right\|_2,
\end{equation}
with equality if and only if
$\Phi_{\rm sym}\propto K_t^{(\mathrm{sym})}$.

The squared norm of the kernel is readily evaluated,
\begin{equation}
\left\|K_t\right\|_2^2
=
\int_{-\infty}^{t} ds'\, e^{(\Gamma_f-\Gamma_e)s'}
\int_{-\infty}^{s'} ds\, e^{\Gamma_e s}
=
\frac{e^{\Gamma_f t}}{\Gamma_e\Gamma_f}.
\end{equation}
Since the supports of $K_t(s^{\prime},s)$ and $K_t(s,s^{\prime})$ are disjoint up to a set of measure zero, one finds
\begin{equation}
\left\|K_t^{(\mathrm{sym})}\right\|_2
=
\frac{1}{\sqrt{2}}\left\|K_t\right\|_2.
\end{equation}
The symmetric two-photon wave function maximizing
$\left|\mathcal I\right|^2$ under the normalization condition (\ref{eq:norm}) is therefore given by
\begin{align}
\Phi_{\rm sym}^{\star}(s^{\prime},s)
=
\frac{\sqrt{\Gamma_e\Gamma_f}\,e^{-\Gamma_f t/2}}{2}
\Big[
& e^{A s^{\prime}} e^{B s}\,
\Theta(t-s')\Theta(s'-s)
\nonumber\\
+
& e^{A s} e^{B s^{\prime}}\,
\Theta(t-s)\Theta(s-s^{\prime})
\Big].
\label{eq: optimalPhi}
\end{align}
For this choice of $\Phi_{\rm sym}$, the functional attains its maximal value and one obtains
\begin{equation}
4\Gamma_e\Gamma_f e^{-\Gamma_f t}
\left|\mathcal I[\Phi_{\rm sym}^{\star}]\right|^2
=
1.
\end{equation}

\section{Maxima of marginal distribution of the optimal state in the frequency domain}\label{Appen: marg_distr_maxima}

Here we consider the problem of determining the values of the local maxima of the marginal distribution (\ref{eq: optimal_state_marg}). An explicit analytical expression for the maxima cannot be obtained in this case, as it would require finding the roots of a fifth-degree polynomial. Since no general analytical solution exists for quintic equations, the problem was instead solved numerically. Further details can be found in Ref. \cite{SupMat1}. Here we consider only two limiting cases, namely, when $\Gamma_{e}\ll\Gamma_{f}$ and $\Gamma_{f}\ll \Gamma_{e}$. Interestingly, analytical expressions for the local maxima can be obtained for these limiting relations between the lifetimes of the excited states.

To solve the problem for $\Gamma_{e}\ll\Gamma_{f}$, it is convenient to introduce the variables 
$r=\Gamma_{e}/\Gamma_{f}$, $d=\delta_{a}/\Gamma_{f}$, and $x = (\omega-\omega_{eg})/\Gamma_{f}$. Determining the maxima then requires finding the roots of the polynomial,
\begin{align}
p(x) = &- 2x^5+ 2dx^4  + \left[- 4rd^2 - 4r(r + 1)\left(r + \frac{1}{4}\right)\right]x^3 \nonumber\\&+ \left[6d\left(rd^2 + r(r + 1)\left(r + \frac{1}{4}\right)\right) - \frac{dr^2}{2}\right]x^2\nonumber\\ &+ \bigg[\frac{1}{2}\left(r^2\left(\frac{(r + 1)^2}{4} + d^2\right)\right) \nonumber\\&- \left(rd^2 + r(r + 1)\left(r + \frac{1}{4}\right)\right)\nonumber\\
&\times\left(\frac{(r + 1)^2}{2} + 2d^2 + \frac{r^2}{2}\right)\bigg]x \nonumber\\&+ \frac{1}{2}dr^2\left[rd^2 + r(r + 1)\left(r + \frac{1}{4}\right)\right].
 \end{align}
In the limit $r\to 0$, the problem reduces to solving an equation of the form
\begin{equation}
x^{4}(-x+d)=0,
\end{equation}
which has two solutions, $x=0$ and $x=d$, which results in $\omega = \omega_{eg}$ and $\omega=\omega_{fe}$.

The solutions for $\Gamma_{f}\ll \Gamma_{e}$ can be determined by introducing  $R=\Gamma_{f}/\Gamma_{e}$, $D=\delta_{a}/\Gamma_{e}$, and $y = (\omega-\omega_{eg})/\Gamma_{e}$, and by formulating the problem of determining the local maxima in terms of a polynomial equation,
\begin{align}
p(y)&=- 2Ry^5 + 2DRy^4 \nonumber
- ( R^2 +5R + 4D^2 +4)y^3 \nonumber\\&
+ \frac{1}{2}D(12D^2 + 3R^2 + 14R + 12)y^2\nonumber\\ 
&+ \bigg[\frac{1}{2}R\left(\frac{(R + 1)^2}{4} + D^2\right) \nonumber\\&- \left(D^2 + \left(\frac{R}{4} + 1\right)(R + 1)\right)\nonumber\\
&\times\left(\frac{(R + 1)^2}{2} + 2D^2 + 1/2\right)\bigg]y \nonumber\\
&+ \frac{1}{2}D\left[D^2 + \left(\frac{R}{4} + 1\right)(R + 1)\right].
\end{align}
One can easily check that for $R\to 0$, the polynomial reduces to the form 
\begin{align}
p(y)&= - 4(D^2 + 1)y^3 + 6D(D^2 + 1)y^2 \nonumber\\
&- (D^2 + 1)(2D^2 + 1)y  -\frac{D(D^2 + 1)}{2},
\end{align}
which for $\delta_{a}<\Gamma_{e}$ has a single root $y=D/2$ corresponding to $\omega=\omega_{fg}/2$.

\section{Derivation of the sum-frequency distribution of the optimal state, Eq.~\eqref{Eq: opimum marignal plus}}\label{Appendix: derivation of distribution of sum}

We derive the distribution of the frequency sum
$p_+(\omega_{+})$ by integrating the joint spectral
density ~\eqref{Eq: optimum spectral prob} over $\omega_1$ and $\omega_2$ with fixed sum $\omega_1 + \omega_2$, i.e.,
\begin{equation}
p_{+}(\omega_{+})
=
\int_{-\infty}^{\infty}
d\omega_{1}
\int_{-\infty}^{\infty}
d\omega_{2}\,
p(\omega_{1},\omega_{2})
\,\delta\!\left(
\omega_{+}-\omega_{1}-\omega_{2}
\right).
\end{equation}
We introduce
\begin{equation}
 y = \omega_1 - \omega_{eg},
\end{equation}
so that $\omega_1 = y+\omega_{eg}$, $\omega_2 = \omega_{+}-y-\omega_{eg}$,
and the Jacobian of the transformation equals unity.
Substituting into ~\eqref{Eq: optimum spectral prob} and integrating over $y$ gives
\begin{equation}
  p_+(\omega_+) = \frac{\Gamma_e\Gamma_f}{8\pi^2}
  \frac{(\omega_{+}-2\omega_{eg})^2+\Gamma_e^2}{(\omega_{+}-\omega_{fg})^2+\tfrac{\Gamma_f^2}{4}}
  \quad I\,,
\end{equation}
where the integral
\begin{equation}
  {I} = \int_{-\infty}^{+\infty} dy\;
  \frac{1}{\bigl[y^2+\frac{\Gamma_e^2}{4}\bigr]
            \bigl[(\omega_{+}-y-2\omega_{eg})^2+\frac{\Gamma_e^2}{4}\bigr]}
\end{equation}
is evaluated by the residue theorem, closing the contour
in the upper half plane.  The poles in the upper
half plane are $y_1 = i\Gamma_e/2$ and
$y_2 = \omega_{+}-2\omega_{eg}+i\Gamma_e/2$.
Computing the two residues and summing yields
\begin{equation}
  {I} = \frac{4\pi}{\Gamma_e}
  \frac{1}{(\omega_{+}-2\omega_{eg})^2+\Gamma_e^2}.
\end{equation}
Substituting back and noting that
$(\omega_{+}-2\omega_{eg})^2+\Gamma_e^2$ cancels with the numerator factor,
one obtains
\begin{equation}
  p_+(\omega_{+})
  = \frac{\Gamma_f}{2\pi}
    \frac{1}{(\omega_{+}-\omega_{fg})^2
              +\frac{\Gamma_f^2}{4}},
\end{equation}
which is Eq.~\eqref{Eq: opimum marignal plus}.

\section{Set of equations for coherent excitation}\label{Appendix: Coherent State}

In this appendix, we present the equations for describing the dynamics of the atomic system interacting with coherent light in the basis of $\ket{g}$, $\ket{e}$, and $\ket{f}$ states. The reduces dynamics of the system is then given by the master equation (\ref{eq: master_equation}), introduced in Sec.~\ref{Section: Coherent state}. Working in the rotation frame defined by (\ref{eq: rotating_frame}) and assuming that $\alpha(t)$ is real without losing generality, we obtain the following differential equations for the elements of the density matrix: 
\begin{widetext}
    
\begin{align}
\dot{\rho}_{ff}(t) &= -\Gamma_{f} \rho_{ff}(t) - \sqrt{\overline{n}}\alpha_{0}(t) \sqrt{\Gamma_{f}} \left(  \rho_{ef}(t) +\rho_{fe}(t) \right), \\[6pt]\label{eq: master_eq1}
\dot{\rho}_{ef}(t) &= \left( i\Delta_2 - \frac{\Gamma_{e} + \Gamma_{f}}{2} \right) \rho_{ef}(t)+\sqrt{\overline{n}}\alpha_{0}(t) \sqrt{\Gamma_{f}} \left( \rho_{ff}(t) - \rho_{ee}(t) \right) - \sqrt{\overline{n}}\alpha_{0}(t) \sqrt{\Gamma_{e}} \rho_{gf}(t) 
, \\[6pt]
\dot{\rho}_{gf}(t) &= \left( i(\Delta_1 + \Delta_2) - \frac{ \Gamma_{f}}{2} \right) \rho_{gf}(t) - \sqrt{\overline{n}}\alpha_{0}(t) \sqrt{\Gamma_{f}} \rho_{ge}(t) + \sqrt{\overline{n}}\alpha_{0}(t) \sqrt{\Gamma_{e}} \rho_{ef}(t) 
, \\[6pt]
\dot{\rho}_{ge}(t) &= \left( i\Delta_1 - \frac{\Gamma_{e}}{2} \right) \rho_{ge}(t) +\sqrt{\overline{n}}\alpha_{0}(t) \sqrt{\Gamma_{e}} \left( \rho_{ee}(t) - \rho_{gg}(t) \right) 
+ \sqrt{\overline{n}}\alpha_{0}(t) \sqrt{\Gamma_{f}} \rho_{gf}(t) + \sqrt{\Gamma_{e} \Gamma_{f}} \rho_{ef}(t), \\[6pt]
\dot{\rho}_{ee}(t) &= - \Gamma_{e} \rho_{ee}(t) +\Gamma_{f} \rho_{ff}(t) 
- \sqrt{\overline{n}}\alpha_{0}(t) \sqrt{\Gamma_{e}} \left( \rho_{eg}(t) + \rho_{ge}(t) \right) 
+ \sqrt{\overline{n}}\alpha_{0} \sqrt{\Gamma_{f}} \left( \rho_{ef}(t) + \rho_{fe}(t) \right), \\[6pt]
\dot{\rho}_{gg}(t) &= \Gamma_{e} \rho_{ee}(t) + \sqrt{\overline{n}}\alpha_{0}(t) \sqrt{\Gamma_{e}} \left( \rho_{ge}(t) + \rho_{eg}(t) \right),\label{eq: master_eq2}
\end{align}
\end{widetext}
where $\rho_{ef}(t) = (\rho_{fe}(t))^{\dagger}$ and so on. We assume that the atom initially is in the ground state, so $\rho_{gg}(t_0) = 1$ and all other elements of the density matrix are equal to zero at the moment when the interaction begins.

\clearpage

\begin{widetext}

\begin{center}
\bf{Supplemental Material for ``Optimization of two-photon excitation by indistinguishable photons in a three-level atom''}

\end{center}

We present here the parameter values that determine the conditions for optimal excitation of a three-level atom for the states of light discussed in the main text. The excitation conditions, together with references to the corresponding states, are provided in the captions.
%Here we discuss and present additional results that support the conclusions of the main text.

%\section{Optimal parameters}
\begin{figure}[h]
    \centering
    \includegraphics[width=0.4\linewidth]{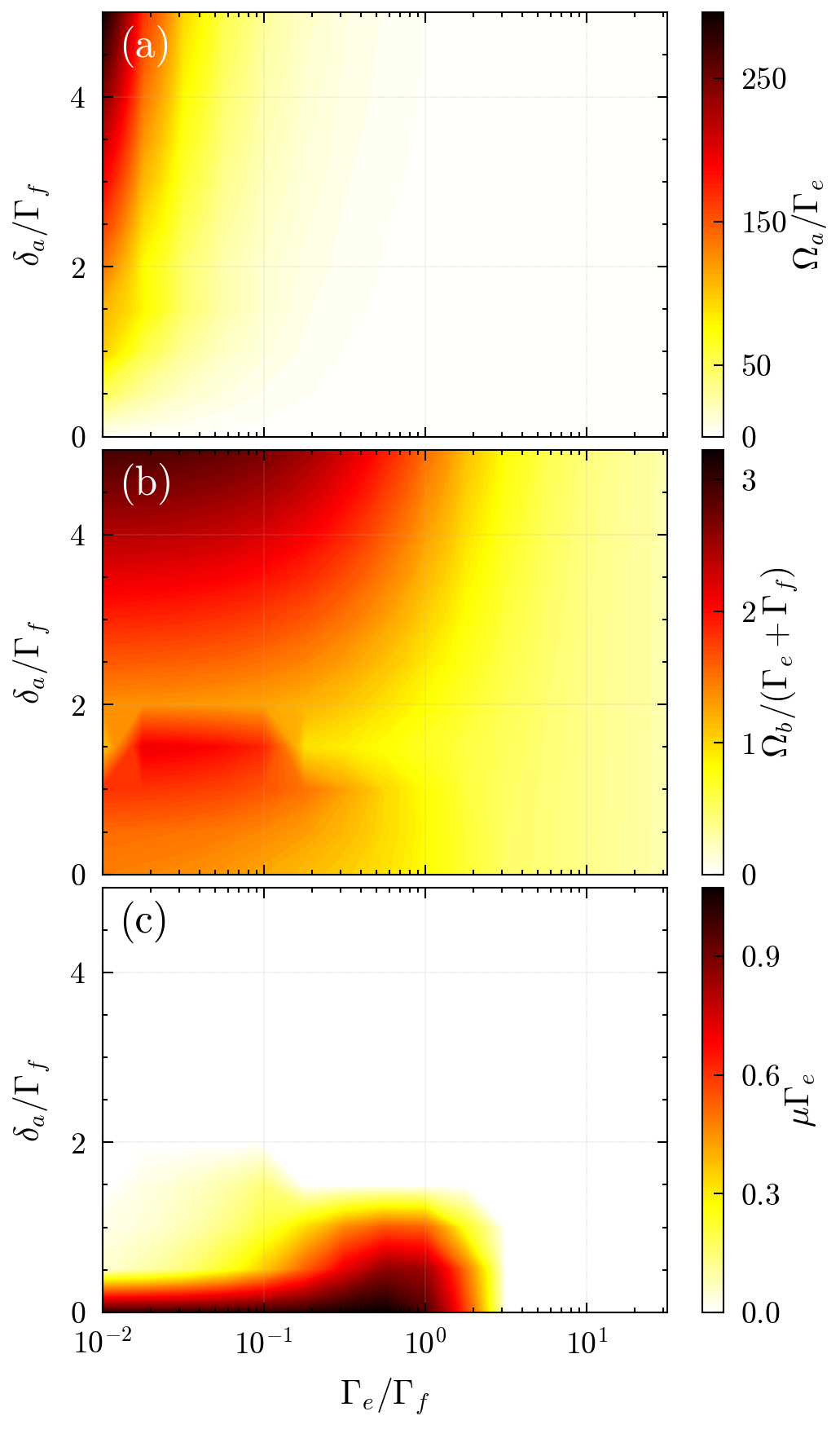}
    \caption{ Values of the corresponding parameters providing maximum excitation probability for the state of light defined by Eq. 29 in the main text, with fixed $\delta_f = 0$ under the double resonance condition. The panels display the optimized values of: (a) $\Omega_a/\Gamma_e$, (b) $\Omega_b/(\Gamma_e + \Gamma_f)$, and (c) $\mu\Gamma_e$  as functions of of the ratio  $\Gamma_e/\Gamma_f$ and the  detuning $\delta_a/\Gamma_f$. All panels are invariant under the transformation $\delta_a/\Gamma_f \to -\delta_a/\Gamma_f$.}
    \label{fig: HeatMap_params_UnEnt_deltaF0}
\end{figure}

\begin{figure}[h]
    \centering
    \includegraphics[width=1\linewidth]{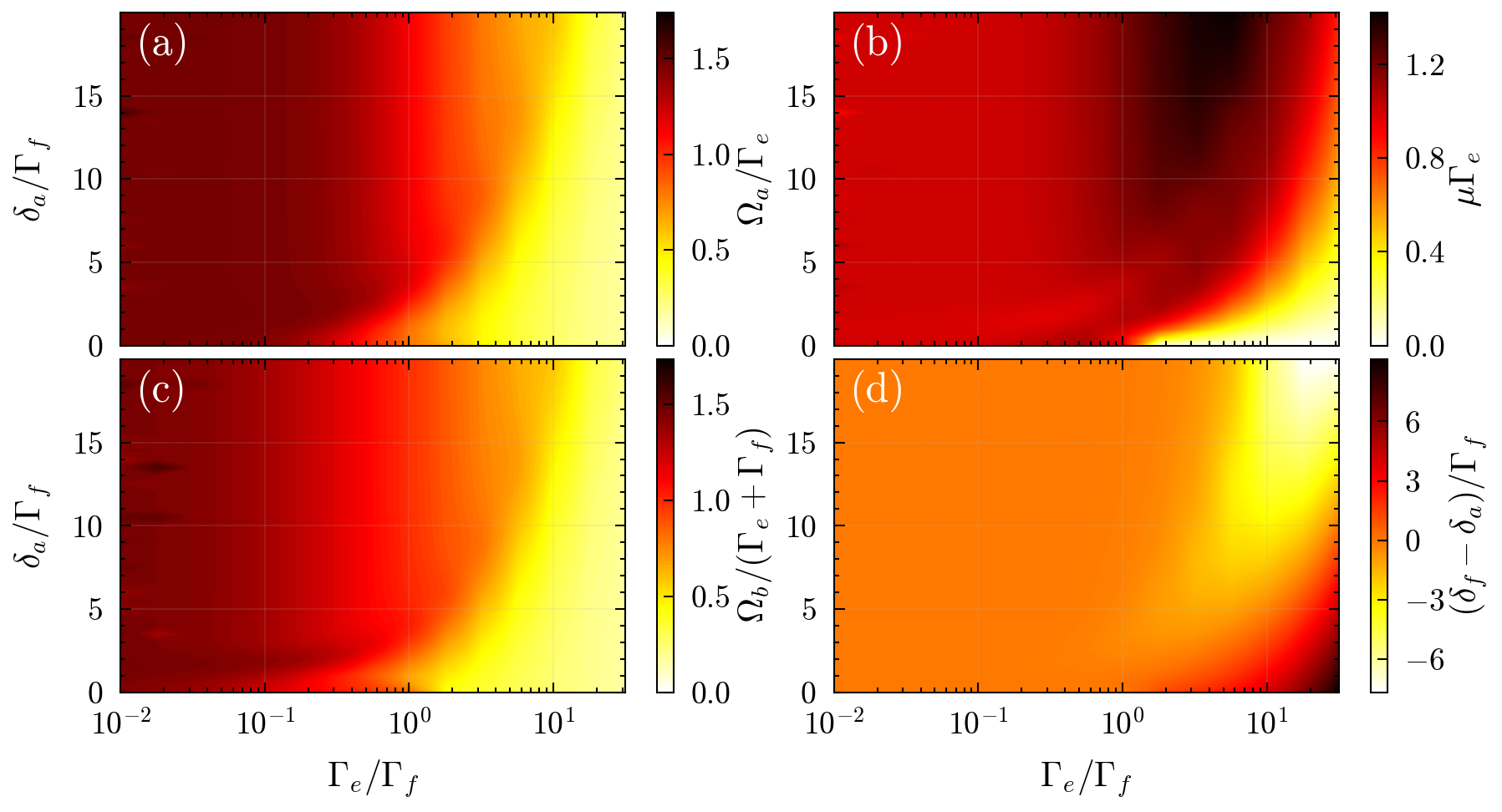}
    \caption{Values of the corresponding parameters providing maximum excitation probability for the state of light defined by Eq. 29 in the main text under the resonance condition. The panels display the optimized values of: (a) $\Omega_a/\Gamma_e$, (b) $\mu\Gamma_e$, (c) $\Omega_b/(\Gamma_e + \Gamma_f)$, and (d) $(\delta_f - \delta_a)/\Gamma_f$ as functions  of the ratio  $\Gamma_e/\Gamma_f$ and the detuning $\delta_a/\Gamma_f$. It can be easily shown that panels (a), (b), and (c) are invariant under the transformation $\delta_a/\Gamma_f \to -\delta_a/\Gamma_f$. }
    \label{fig: HeatMap_params_UnEnt}
\end{figure}

\begin{figure}[h]
    \centering
    \includegraphics[width=0.4\linewidth]{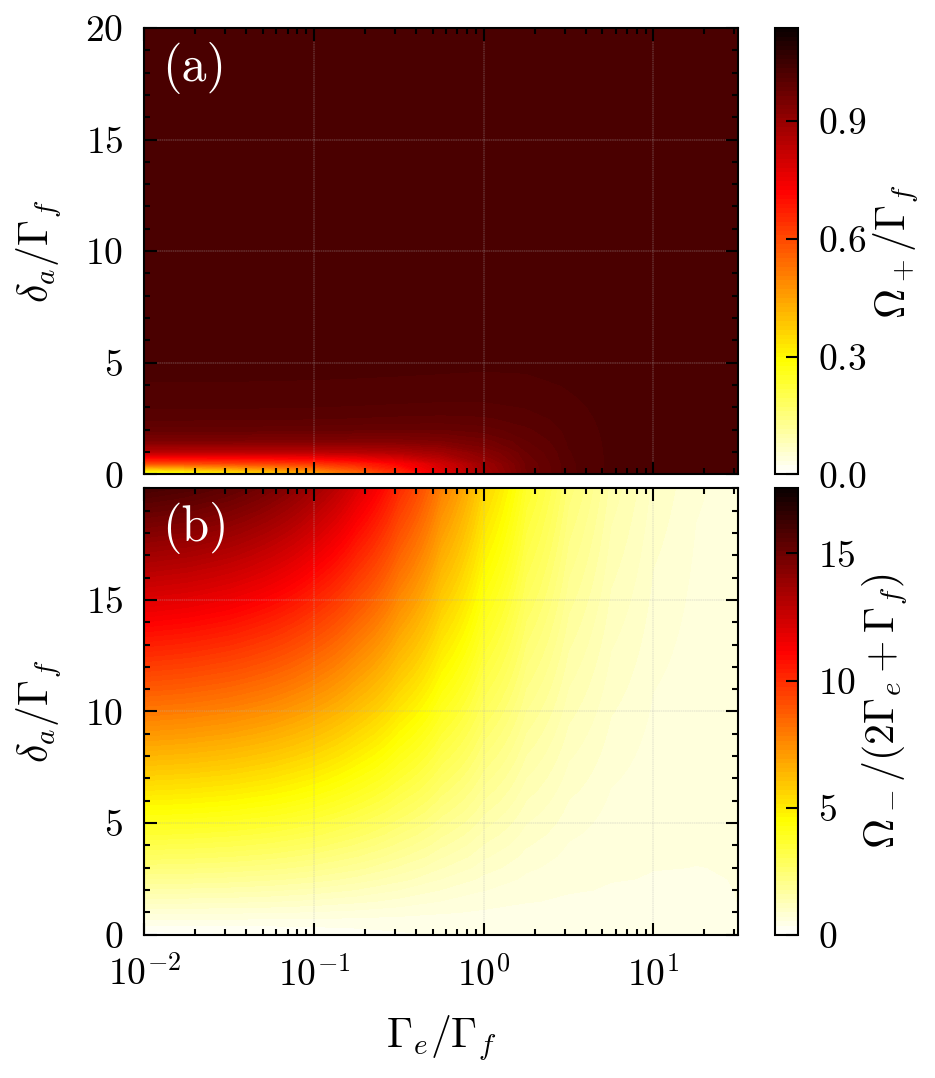}
   \caption{Values of the corresponding parameters providing maximum excitation probability for the state of light defined by Eq. 39 in the main text, with fixed $\delta_f = 0$  under the double resonance condition. The panels display the optimized values of: (a) $\Omega_+/\Gamma_f$, and (b) $\Omega_-/(2\Gamma_e + \Gamma_f)$ as functions of the ratio $\Gamma_e/\Gamma_f$ and the detuning $\delta_a/\Gamma_f$. Both panels are invariant under the change of the sign of $\delta_a/\Gamma_f$.}
    \label{fig: HeatMap_params_Entangled_deltaF0}
\end{figure}

\begin{figure}[h]
    \centering
    \includegraphics[width=0.4\linewidth]{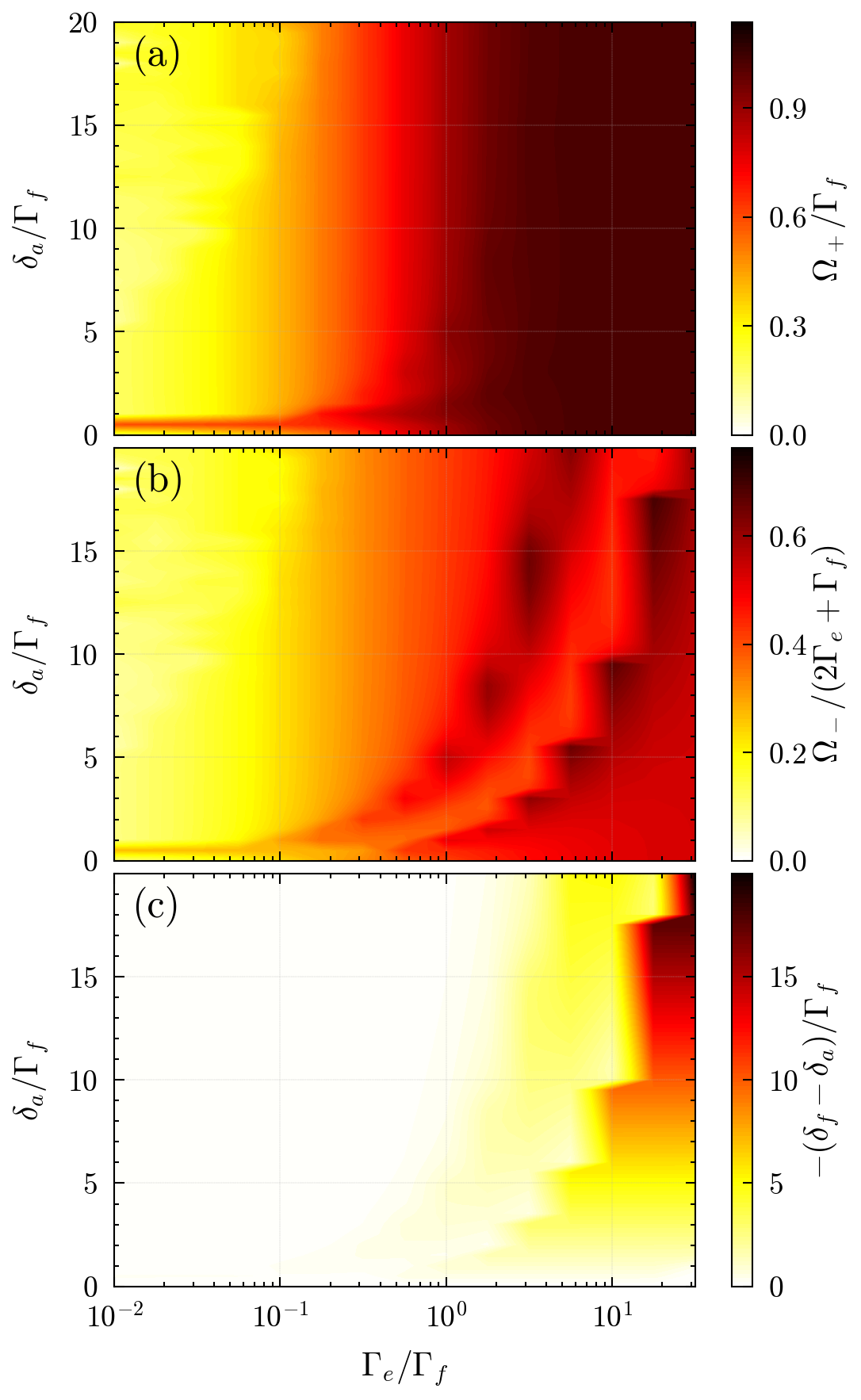}
   \caption{Values of the corresponding parameters providing maximum excitation probability for the state of light defined by Eq. 39 in the main text under the double resonance condition. The panels display the optimized values of: (a) $\Omega_+/\Gamma_f$, (b) $\Omega_-/(2\Gamma_e + \Gamma_f)$, and (c) $-(\delta_f - \delta_a)/\Gamma_f$ as functions of the ratio $\Gamma_e/\Gamma_f$ and the detuning $\delta_a/\Gamma_f$. Panels (a) and (b) are invariant under the transformation $\delta_a/\Gamma_f \to -\delta_a/\Gamma_f$.  }
    \label{fig: HeatMap_params_Entangled}
\end{figure}

\begin{figure}[h]
    \centering
    \includegraphics[width=0.4\linewidth]{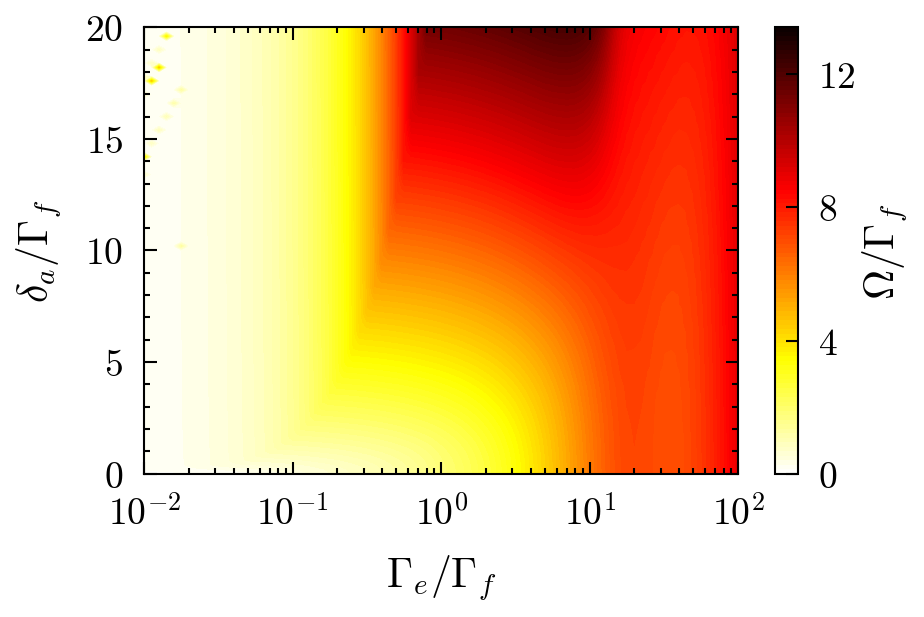}
    \caption{Values of the parameters $\Omega/\Gamma_f$ providing maximum excitation probability for the coherent state with $\Bar{n} = 2$, shown as a function of the ratio $\Gamma_e/\Gamma_f$ and the detuning $\delta_a/\Gamma_f$ under double resonance condition. The plot is invariant under the transformation $\delta_a/\Gamma_f \to -\delta_a/\Gamma_f$.}
    \label{fig: HeatMAp_Params_coherent}
\end{figure}

\end{widetext}

\end{document}